\documentclass[12pt]{article}
\usepackage{enumerate}
\usepackage{natbib}
\usepackage{amsmath, amssymb, bbm, setspace, mathrsfs, color, listings, graphics, graphicx, multicol, dcolumn, dsfont, subcaption, float, mathabx}
\usepackage[font=footnotesize,labelfont=bf,skip=5pt]{caption}
\usepackage{multirow}
\usepackage{hyperref}
\usepackage{breakurl}
\usepackage[titletoc, toc, page]{appendix}

\DeclareMathOperator*{\Y}{\mathbf{Y}}
\newcommand{\paren}[1]{\!\left( #1 \right)}
\newcommand{\set}[1]{\left\lbrace #1 \right\rbrace}

\newcommand{\N}[2]{\text{N}\left(#1, #2\right)}

\newcommand{\blind}{1}

\begin{document}

\def\spacingset#1{\renewcommand{\baselinestretch}%
{#1}\small\normalsize} \spacingset{1}

%%%%%%%%%%%%%%%%%%%%%%%%%%%%%%%%%%%%%%%%%%%%%%%%%%%%%%%%%%%%%%%%%%%%%%%%%%%%%%

\if1\blind
{
  \title{\bf Modeling a Nonlinear Biophysical Trend Followed by Long-Memory Equilibrium with Unknown Change Point}
  \author{Wenyu Zhang \\
  Department of Statistics and Data Science, \\ Cornell University \\ 
  and \\ 
  Maryclare Griffin \\
  Department of Mathematics and Statistics, \\ University of Massachusetts Amherst \\
  and \\
  David S.\ Matteson \thanks{The authors gratefully acknowledge financial support from the Cornell University Institute of Biotechnology, the New York State Foundation of Science, Technology and Innovation (NYSTAR), a Xerox PARC Faculty Research Award, National Science Foundation Awards 1455172, 1934985, 1940124, and 1940276, USAID, and Cornell University Atkinson Center for a Sustainable Future.} \\
  Department of Statistics and Data Science, \\ Cornell University }
  \date{}
  \maketitle
} \fi

\if0\blind
{
  \bigskip
  \bigskip
  \bigskip
  \begin{center}
    {\LARGE\bf Modeling a Nonlinear Biophysical Trend Followed by Long-Memory Equilibrium with Unknown Change Point}
\end{center}
  \medskip
} \fi

\bigskip
\begin{abstract}
Measurements of many biological processes are characterized by an initial trend period followed by an equilibrium period. Scientists may wish to quantify features of the two periods, as well as the timing of the change point. 
Specifically, we are motivated by problems in the study of electrical cell-substrate impedance sensing (ECIS) data.
ECIS is a popular new technology which measures cell behavior non-invasively. 
Previous studies using ECIS data have found that different cell types can be classified by their equilibrium behavior. 
However, it can be challenging to identify when equilibrium has been reached, and to quantify the relevant features of cells' equilibrium behavior.
In this paper, we assume that measurements during the trend period are independent deviations from a smooth nonlinear function of time, and that measurements during the equilibrium period are characterized by a simple long memory model. We propose a method to simultaneously estimate the parameters of the trend and equilibrium processes and locate the change point between the two. We find that this method performs well in simulations and in practice. When applied to ECIS data, it produces estimates of change points and measures of cell equilibrium behavior which offer improved classification of infected and uninfected cells. Code for the implementation is publicly available\footnote{https://github.com/zwenyu/T2CD}.
\end{abstract}

\noindent%
{\it Keywords:}  Applied biophysics; change point analysis; fractionally integrated process; long memory; time series.
\vfill

\newpage
\spacingset{1.5}

\section{Introduction}
\label{sec: intro}

We propose a model for time-series data that is characterized by two consecutive regimes, which correspond to a highly nonstationary and nonlinear trend period and a stable, equilibrium period.
Often, researchers are interested in estimating the features of each regime, as well as the timing of the transition or change point between the two.

We are motivated by the problem of detecting contamination of mammalian cell cultures by mycoplasma using electric cell-substrate impedance sensing (ECIS) data.
Contamination of mammalian cell cultures is pervasive, costly, and can be challenging to detect \citep{gustavsson2019}. 
Specifically, contamination by mycoplasma is especially prevalent, occurring in up to 20\% of cell cultures, while also expensive and time consuming to detect.
As a result, there is a pressing need for the development of additional methods for detecting contamination by mycoplasma.

ECIS is a relatively new non-invasive method used to study cell attachment, growth, morphology, function and motility \citep{ecis2019}. 
ECIS measurements have been used in numerous cell biology studies, from cancer biology and cytotoxicity \citep{opp2009,hong2011}.
Because ECIS measurements have been used to differentiate between cancerous and noncancerous cells and to classify cell lines \citep{lovelady2007,gelsinger2017}, it is hypothesized that they may also be used to identify cell cultures contaminated by mycoplasma.

\begin{figure}
\centering
\includegraphics[width=0.7\linewidth]{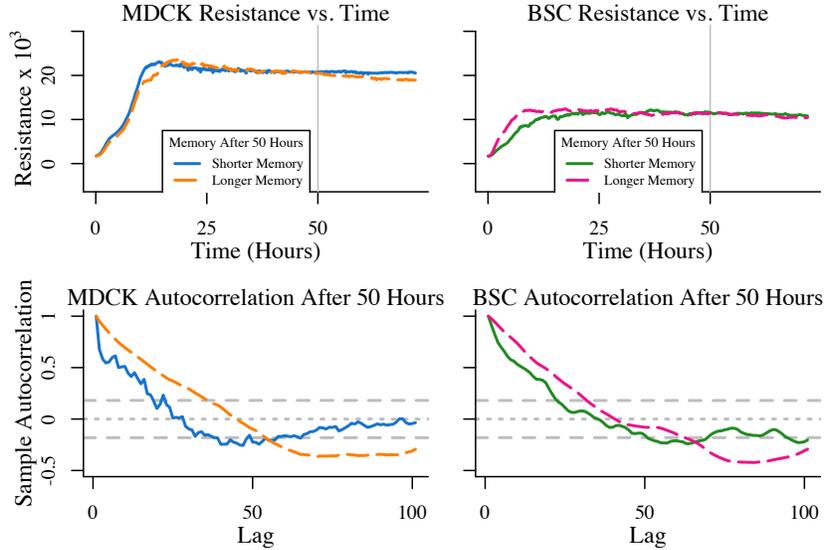}

\caption{The first row shows examples of resistance measurements at 4000 hertz for MDCK and BSC cell line samples cultivated in gel. For each cell line, one example of resistance measurements displaying shorter memory and one example of resistance measurements displaying longer memory are selected. A gray vertical line at 50 hours is provided to indicate a conservative estimate of the onset of the equilibrium regime. The second row shows the corresponding sample autocorrelations for resistance measurements after 50 hours. Approximate 95\% intervals for sample autocorrelations in the absence of dependence across time during the equilibrium regime are given in gray.}
\label{fig: norinf}
\end{figure}

ECIS measurements are obtained by growing cells in a well on top of small gold-film electrodes, between which alternating current is applied and electrical impedance is measured.  
As cells grow, they cover the electrode and resistance, a component of impedence, increases.
Eventually, the cells fill the well and growth ceases. In some cases, cell death occurs due to overcrowding, causing a small drop in resistance measurements after the peak.
After this point, an equilibrium period begins. Resistance fluctuations during equilibrium are caused by cell micromotion.
The equilibrium period is sometimes called confluence in the ECIS literature, and it continues until the cells exhaust their resources and begin to die. 
The first row of Figure \ref{fig: norinf} shows a subset of resistance measurements for two different cell types, Madin-Darby Canine Kidney (MDCK) cells and epithelial cells of African green monkey kidney origin (BSC-1 cells). All show a nonlinear trend period followed by an equilibrium period, with a more visually obvious change point present in the MDCK cells. A color version of the figures can be found in the electronic version of the article.

Equilibrium measurements are especially informative. They are believed to be less sensitive to initial conditions than features of the trend period, and can be characterized parsimoniously by a very simple three-parameter long memory time series model \citep{lovelady2007, tarantola2010}. 
Equilibrium measurements display long-range dependence; the correlations between successive measurements decay very slowly over time. 
The second row of Figure \ref{fig: norinf} shows sample autocorrelation functions for the subset of resistance measurements shown in the first row resistance measurements after 50 hours, at which point equilibrium has been achieved for all four selected time series. The selected time series were chosen to show how the rate of  different resistance time series may display shorter or longer memory, i.e. weaker or stronger long-range dependence.

Long-range dependence has also been observed in wind speed and inflation data \citep{haslett1989, jurgen2004}, and can modeled as a Gaussian fractionally integrated (FI) or long-memory process. % Cite  
The FI process has three parameters, an overall mean $\mu$, variance $\sigma^2$, and a scalar long-memory (fractional differencing) parameter $d$ that governs how quickly auto-correlations decay.
Ideally, if these parameters could be estimated well, they could be used to quantify features of the equilibrium regime in the context of ECIS data.

Unfortunately, the long-memory parameter $d$ is notoriously difficult to estimate in finite samples.
Furthermore, the change point from trend to confluence phase, which determines the amount of data available to estimate $d$, is typically not precisely known in practice.
Standard practice is use a fixed time point, e.g., 20 hours, as a conservative estimate of the start of the confluence regime \citep{tarantola2010}. 
This under-utilizes the data, potentially resulting in poorer estimates of the parameters of interest. Furthermore, such a conservative estimate could incorrectly characterize the preceding trend phase. 

This suggests the need for a change point detection method which can identify when the trend phase gives way to confluence, specifically an unsupervised method that can detect the transition from a nonstationary model to a FI model.

To our knowledge, existing methods for change-point detection are not appropriate.
Some existing methods assume a short-memory autoregressive moving average (ARMA), long-memory FI models, or other restrictive parametric models both before and after the change point 
\citep{chen1993,gilles2008,killick12}.
Others assume that measurements between change points are independent or identically distributed, or assume that change points strictly correspond to level shifts or isolated outliers \citep{matteson2014,zhang2019}. 
Alternative methods in the biomedical fields tend to be too domain-specific to apply to the the problem of detecting the change point between the trend and confluence phase in ECIS data \citep{olshen04, nika2014}.

In this paper, we develop a novel method for estimating a change point between a highly nonstationary and nonlinear trend period and a stable, equilibrium period that is characterized by an FI process. We simultaneously obtain estimates of the nonlinear trend function and the FI parameters.
We apply this method to the detection of contamination by M. hominis, a species of mycoplasma, in MDCK cells and BSC-1 cells using ECIS measurements.
The available data consists of four experiments per cell type. 
Each experiment corresponds to ECIS measurements on cells on a single tray of 96 wells obtained over the course of at least 72 hours. Of the 96 wells, 16 are left empty, 32 contain uncontaminated cells and 48 contain cells contaminated by mycoplasma.
In order to mimic lab-to-lab variability in cell culture preparation, wells were prepared using either of two different types of media. Half were prepared using bovine serum albumin (BSA), and half were prepared using gel.
Within an experiment, wells containing the same media and cells with the same contamination status can be thought of as replicates.

In Section~\ref{sec:methodology}, we propose a model which we call Trend-to-Confluence Detector (T2CD) for data which display highly nonstationary and nonlinear trend period followed by a stable, equilibrium period with long-range dependence. 
In Section~\ref{sec:estimation}, we discuss estimation of the parameters of the model introduced in Section~\ref{sec:methodology}. We consider both an exact estimation procedure which we call T2CD-step, as well as an generalized estimation procedure which has greater computational scalability for longer time series which we call T2CD-sigmoid.
We demonstrate the performance of T2CD-step and T2CD-sigmoid in simulations in Section~\ref{sec:simulations}. We apply T2CD-step and T2CD-sigmoid to the ECIS data shown in Figure \ref{fig: norinf} and use the estimated change points and FI parameters to better classify cells by contamination status in Section~\ref{sec:ecis}.

\section{Trend-to-Confluence Detector (T2CD) Model}
\label{sec:methodology}

\subsection{Overview}
\label{subsec:overview}

Let $y_1, y_2, \dots, y_T \in \mathbb{R}$ be a sequence of time-ordered observations at $t = 1, 2, \dots, T$, respectively. We assume that the measurements $y_t$ belong to two successive regimes, a trend regime and an equilibrium or confluent regime. 

Let $\tau$ denote the change point time index. We assume that
  \begin{align}
y_t &= f\paren{t; \mathbf{\beta}} + \eta_t & \quad \text{for $t < \tau$}\\
y_t  &= g\paren{y_1, \dots, y_{t-1}; \mu, d, \tau} + \epsilon_t
& \quad \text{for $t \geq \tau$}
\end{align}
where $\eta_t \sim \N{0}{\text{exp}\{h\paren{t; \mathbf{\theta}}\}}$ and $\epsilon_t \sim \N{0}{\nu^2}$, $\tau_a \leq \tau \leq \tau_b$, and $\tau_a$ and $\tau_b$ are the  prespecified minimum and maximum values of the change point chosen according to a priori knowledge of the change point location. In the absence of a priori information, $\tau_a = 0$ and $\tau_b = T$. The noise terms $\eta_t$ and $\epsilon_t$, which encompass measurement errors and random fluctuations due to continuous cell growth, motility, and death, are assumed to be independent within and across the two regimes.

During the first regime $( t < \tau )$, the measurement at time $t$ will be centered around a trend curve $f\paren{t; \mathbf{\beta}}$ which is a function of time $t$ and fixed but unknown parameters $\mathbf{\beta}$. The noise terms $\{ \eta_t \}$ %$\eta_t \sim \N{0}{\sigma^2_t}$
are possibly heteroscedastic with variance $\text{exp}\{h\paren{t; \mathbf{\theta}\}}$, to reflect different degrees of uncertainty in the measurements when the cell culture undergoes different rates of growth and death. 
During the second (equilibrium) regime $( t \ge \tau )$, the measurement at time $t$ will be centered about a function $g\paren{y_{1}, \dots, y_{t-1}; \mu, d, \tau}$ of previous measurements $y_{1}, \dots, y_{t-1}$ and fixed but unknown parameters $\mu$, and $d$. The noise terms $\epsilon_t \sim \N{0}{\nu^2}$ are homoscedastic with fixed but unknown variance $\nu^2$, since the cell culture is in equilibrium and not undergoing drastic changes. We describe our modeling choices of the two regimes in the following sections.

\subsection{Trend}
\label{subsec:trend}

Resistance measurements in the first regime, or the trend phase, are characterized by a trend of initial steep increase sometimes followed by a slight drop after the peak, as well as heteroscedasticity with higher variance at the stage of rapid cell growth. 
As mentioned above, the trend curve is denoted as $f\paren{t;\beta}$.
Depending on the trend, any appropriate parametric, semi-parametric or nonparametric model can be used to fit the first regime. The exact formulation of the trend curve can depend on the application domain and the choice of the user. 
For the ECIS application that we focus on in this paper, we assume a smooth trend curve. This is in line with visual inspection of real ECIS data in Figure \ref{fig: norinf}, and that cell growth, motility, death and other functions are continuous processes.
We utilize penalized splines \citep{ruppert2003} for their flexibility to capture the ECIS trend phase, since it is highly nonstationary. We similarly use penalized splines in modeling the noise variance. 

We denote the matrix of B-spline basis functions $B_{i,D}(u)$ of degree $D_f$ evaluated on time indices for the trend as $\mathbf{X} = \left(\mathbf{x}_1', \mathbf{x}_2', \dots, \mathbf{x}_{T}' \right) \in \mathbbm{R}^{T \times (Q_f+D_f+1)}$, where $Q_f$ is the number of distinct interior knots.
Similarly, we denote $\mathbf{V} = \left(\mathbf{v}_1', \mathbf{v}_2', \dots, \mathbf{v}_{T}' \right)$ as the $T\times \paren{Q_h + D_h + 1}$ matrix of B-spline basis functions for the log variances of the noise terms.
The model for the first regime takes the form
\begin{align*}
    y_t = \mathbf{x}_{t}'\mathbf{\beta} + \eta_t,
\end{align*}
where $\eta_t \sim N(0, \text{exp}\{ \mathbf{v}_{t}'\mathbf{\theta}\})$.
Let the fitted spline for the trend be $s(t) = \mathbf{x}_t' \beta$.
We impose the smoothness penalty $\lambda_f \int \hat{s}''(u)^2 du = \lambda_f \boldsymbol \beta'\boldsymbol M_f \boldsymbol \beta$ on the spline estimate to prevent overfitting,
where $\lambda_f > 0$ is a scalar that determines the smoothness of the fitted  spline and $\boldsymbol M_f$ is a matrix with elements that are fixed given the matrix of B-spline basis functions $\mathbf{X}$.
An equivalent smoothness penalty $\lambda_h \boldsymbol \theta'\boldsymbol M_h \boldsymbol \theta$ is imposed on the fit for the log variances of the noise terms, where $\lambda_h$ is another smoothness parameter and $\boldsymbol M_h$ is a matrix with elements that are fixed given the matrix of B-spline basis functions $\mathbf{V}$.
 
\subsection{Equilibrium}

Starting at time index $\tau$, measurements are centered about a function $g\paren{y_{1}, \dots, y_{t-1}; \mu, d, \tau}$ of previous measurements $y_{1}, \dots, y_{t-1}$ and fixed but unknown parameters $\mu$, and $d$ that corresponds to the conditional mean function of a fractionally integrated (FI) process: 
\begin{align}\label{eq: fi}   
g\paren{y_1, \dots, y_{t-1}; \mu, d, \tau} = \mu - \sum_{i=1}^{t-1} {d \choose i} (-1)^{i} \paren{y_{t-i} - \mu}\mathbbm{1}_{t-i \geq \tau},
\end{align}

This captures long-range dependence of the measurements in confluence.
The parameter $d$ plays the role of the long memory parameter in a FI model \citep{sowell1992}. The FI model assumes that observed values of a time series $y_t$ satisfy $(1 - B)^d y_t = \epsilon_t$, where $B$ is the differencing operator $B^k y_t = y_{t-k}$ and $\epsilon_t \sim \N 0 \nu^2$. 
Values of $d > 0$ correspond to processes that are said to have long memory, with 
larger values of $d$ indicating more slowly decaying autocorrelations over time.
Specifically, the autocorrelation function $Corr(y_t, y_{t-k})$ exhibits hyperbolic decay: as $k \rightarrow \infty$, $Cor(y_t, y_{t-k}) \rightarrow (\Gamma(1 - d)/\Gamma(d))k^{2d-1}$ \citep{baillie1996}.
When $d < 1$, the FI process is mean reverting, and when $d < 0.5$ the FI process is stationary.

\subsection{An Extension to Multivariate Data}
\label{sec: multivariate}

To accommodate settings where $p$ related time series may be observed contemporaneously, we provide an extension to multivariate time series data $\Y \in \mathbb{R}^{T\times p}$.
We assume that all $p$ time series share a common long-memory parameter $d$, but have their own change point $\tau_j$, trend parameters $\mathbf{\beta}_j$ and $\mathbf{\theta}_j$, and equilibrium mean and variance $\mu_j$ and $\nu^2_j$.
Specifically, we assume
\begin{align}
y_{t,j} &= f\paren{t; \mathbf{\beta}_j} + \eta_{t,j} & \quad \text{for $t < \tau_j$}\\
y_{t,j} &= g\paren{y_{1,j}, \dots, y_{t-1,j}; \mu_j, d, \tau_j} + \epsilon_{t,j} & \quad \text{for $t \geq \tau_j$}
\end{align}
where $\eta_{t,j} \sim \N{0}{\text{exp}\{h\paren{t; \mathbf{\theta}_j}\}}$ and  $\epsilon_{j} \sim \N{0}{\nu^2_j}$.

This is motivated by the ECIS measurements described in Section~\ref{sec: intro}, where the $p$ related time series correspond to wells containing cells of the same type, contamination status, and media in the same experiment which may have varying initial conditions but common equilibrium behavior. We account for varying initial conditions, such as the number of cells deposited, by allowing each well to have its own varying change point $\tau_j$, trend parameters $\mathbf{\beta}_j$ and $\mathbf{\theta}_j$, and equilibrium mean and variance $\mu_j$ and $\nu^2_j$.  A shared long-memory parameter $d$ reflects the cells' common equilibrium or confluence behavior.

\section{Estimation}
\label{sec:estimation}

\subsection{Exact Estimation for Univariate Data: T2CD-step}

First, we introduce a strategy for estimating the T2CD parameters that we call T2CD-step, because it performs a complete search  over the change point location space $[\tau_a, \tau_b]$.
We find the change point $\tau_a \leq \hat{\tau}\leq \tau_b$ which maximizes the penalized log-likelihood:
\begin{align}
\label{eqn: loglik}
    %\ell\left(\boldsymbol y \right) =&
    &
    -\sum_{t = 1}^{\tau - 1} \left(\frac{\boldsymbol v_t'\boldsymbol \theta}{2} +\frac{\left(y_t - \boldsymbol x_t' \boldsymbol \beta\right)^2}{2\text{exp}\left\{\boldsymbol v_t'\boldsymbol \theta\right\}}\right) - \frac{1}{2} \lambda_f \boldsymbol \beta'\boldsymbol M_f \boldsymbol \beta - \frac{1}{2} \lambda_h \boldsymbol \theta'\boldsymbol M_h \boldsymbol \theta + \\ \nonumber
    &\hspace{2cm} -\frac{1}{2\nu^2}\sum_{t = \tau}^{T}  \left(y_t - g\left(y_{1},\dots, y_{t-1}; \mu, d, \tau\right) \right)^2 - \left(\frac{T - \tau + 1}{2}\right)\text{log}\left(\nu^2\right) + constant.
\end{align}

Given a candidate change location, the penalized log-likelihood can be decomposed into one component that involves the values of the time series during the trend regime $y_1,\dots, y_{\tau-1}$ and the parameters of the trend regime, $\boldsymbol \beta$ and $\boldsymbol \theta$, and the smoothness parameters, $\lambda_f$ and $\lambda_h$ and another component that involves the values of the time series during the equilibrium period $y_{\tau},\dots, y_T$  and the parameters of the equilibrium period, $\mu$, $d$, and $\nu^2$. 
It follows that the parameters of the trend and equilibrium regime can be estimated simultaneously from the trend and equilibrium data, respectively. 

We estimate the parameters of the trend regime using an iterative Feasible Generalized Least Squares (FGLS) procedure \citep{kuan2004} to estimate the spline coefficients $\mathbf{\beta}$ and $\mathbf{\theta}$, with $\lambda_f$ and $\lambda_h$ chosen according to leave-one-out cross validation as implemented in \texttt{smooth.spline} in \texttt{R} \citep{R} which selects the smoothness penalties by golden-section search. A more detailed explanation of the FGLS procedure is provided in Web Appendix A. 

We estimate the parameters of the equilibrium regime by computing two estimates of the long memory parameter $d$, one by maximizing \eqref{eqn: loglik} over the range $d\in(-0.5, 0.5)$ and another by maximizing \eqref{eqn: loglik} over the range $d \in(0.5, 1.5)$. The conditional mean function $g\left(y_{1},\dots, y_{t-1}; \mu, d, \tau\right)$ takes a first difference when $d\in(0.5, 1.5)$. As a result, discontinuities can occur at $d = 0.5$. For this reason, we choose the estimate of the the long-memory parameter parameter $d$ that is further from the boundary of $0.5$.

\subsection{Generalized Estimation for Univariate Data: T2CD-sigmoid}

In practice, maximizing the penalized log-likelihood \eqref{eqn: loglik} can be prohibitively computationally demanding and time consuming if there are many candidate change points, as is the case when the observed time series is long. Accordingly, we introduce a generalization to the estimation procedure that we call T2CD-sigmoid. 
Let $w\paren{t;\boldsymbol \alpha}$ denote a transition function that takes on values in the interval $\left[0, 1\right]$, then we can define the mean function in the second regime $g\paren{y_1,\dots, y_{t-1}; \mu, d, \tau}$ as defined in \eqref{eq: fi} as a special case of
\begin{align} \label{eq: figen}
\mu - \sum_{i=1}^{t-1} {d \choose i} (-1)^{i} \paren{y_{t-i} - \mu}w\paren{t-i; \boldsymbol \alpha},
\end{align}
where  $w\paren{t;\boldsymbol \alpha}$ has a single parameter $\alpha$ that corresponds to the change point $\tau$ and 
$w\paren{t; \alpha}=\mathbbm{1}_{t-i \geq \alpha}$ that takes the form of a step function.
This suggests that an alternative approach would be to replace the discrete step transition function $\mathbbm{1}_{t-i \geq \tau}$ 
with a continuous sigmoid transition function $w\paren{t; \boldsymbol \alpha} = \paren{1 + \text{exp}\{-\alpha_0 - \alpha_1 t\}}^{-1}$,
which is parameterized by a pair of  real-valued parameters $\boldsymbol \alpha = \set{\alpha_0, \alpha_1}$.
We denote the corresponding second regime mean function as
\begin{align*}
    \tilde{g}\paren{y_1, \dots, y_{t-1}; \mu, d, \alpha_0, \alpha_1} =  \mu - 
    \sum_{i=1}^{t-1} {d \choose i} (-1)^{i} \paren{y_{t-i} - \mu}\paren{1 + \text{exp}\{-\alpha_0 - \alpha_1 (t-i)\}}^{-1}.
\end{align*}
The parameters $\alpha_0$ and $\alpha_1$ determine the timing of the transition from trend to equilibrium phase,
which corresponds to the inflection point of the transition function $\paren{1 + \text{exp}\{-\alpha_0 - \alpha_1 t\}}^{-1}$. 
The change point as estimated as when the transition function is at $0.5$, that is, $\hat{\tau} = -\frac{\hat{\alpha}_0}{\hat{\alpha}_1}$.
The timing of the transition can be constrained to the interval $\left[\tau_a, \tau_b\right]$ by adding a penalty $C(w(\tau_b;\alpha_0, \alpha_1) - w(\tau_a;\alpha_0, \alpha_1))$ with fixed penalty parameter $C > 0$ to the objective function.

Using a smooth transition function can offer computational speed-ups because the log-likelihood can be differentiated with respect to the parameters that determine the timing of the transition, $\alpha_0$ and $\alpha_1$, and accordingly does not require an exhaustive search over all candidate change points. 

The penalized log-likelihood used for T2CD-sigmoid is 
\begin{align}
\label{eqn: loglikapp}
    &    -\sum_{t = 1}^{T} 
    \paren{1-w(t;\alpha_0, \alpha_1)} \left(\frac{\boldsymbol v_t'\boldsymbol \theta}{2} +\frac{\left(y_t - \boldsymbol x_t' \boldsymbol \beta\right)^2}{2\text{exp}\left\{\boldsymbol v_t'\boldsymbol \theta\right\}}\right)
    - \frac{1}{2} \lambda_f \boldsymbol \beta'\boldsymbol M_f \boldsymbol \beta - \frac{1}{2} \lambda_h \boldsymbol \theta'\boldsymbol M_h \boldsymbol \theta +  \\ \nonumber
    &\hspace{2cm} -\frac{1}{2\nu^2}\sum_{t = 1}^{T} w(t;\alpha_0, \alpha_1) \left(y_t - \tilde{g}\left(y_{1},\dots, y_{t-1}; \mu, d,\alpha_0, \alpha_1\right) \right)^2 + \nonumber \\    
    &\hspace{2cm}
    - \frac{1}{2} \sum_{t = 1}^{T} 
    w(t;\alpha_0, \alpha_1) \text{log}\left(\nu^2\right) +  C\paren{w(\tau_b;\alpha_0, \alpha_1) - w(\tau_a;\alpha_0, \alpha_1)} + constant\nonumber
\end{align}
where $C > 0$ is a constant that can be set to ensure that the inflection point of the smooth transition function occurs between $\tau_a$ and $\tau_b$. The intuition for weighing the log-likelihood can also be found in tempered likelihoods for Bayesian inference where a model likelihood is down-weighted if model misspecification is suspected \citep{Thomas2019DiagnosingMM}.

The penalized log-likelihood used by T2CD-sigmoid cannot be decomposed into two components, one of which involves the values of the time series during the trend period and corresponding parameters  and another component that involves the values of the time series during the equilibrium period and the corresponding parameters.
Fortunately, B-spline bases are flexible enough to fit local trends. Accordingly, the first step of T2CD-sigmoid is to estimate the trend regime parameters from the entire time series by maximizing the penalized log likelihood 
\begin{align}
\label{eqn: trendonly}
    &    -\sum_{t = 1}^{T} \left(\frac{\boldsymbol v_t'\boldsymbol \theta}{2} +\frac{\left(y_t - \boldsymbol x_t' \boldsymbol \beta\right)^2}{2\text{exp}\left\{\boldsymbol v_t'\boldsymbol \theta\right\}}\right)  - \frac{1}{2} \lambda_f \boldsymbol \beta'\boldsymbol M_f \boldsymbol \beta - \frac{1}{2} \lambda_h \boldsymbol \theta'\boldsymbol M_h \boldsymbol \theta. 
\end{align}
Again, we use an iterative Feasible Generalized Least Squares (FGLS) procedure \citep{kuan2004} to estimate the spline coefficients $\mathbf{\beta}$ and $\mathbf{\theta}$, with $\lambda_f$ and $\lambda_h$ chosen according to leave-one-out cross validation as implemented in \texttt{smooth.spline} in \texttt{R} \citep{R}.
Via simulations provided in Web Appendix A, we  show that estimates of the trend regime parameters obtained from this procedure are comparable to estimates of the trend regime parameters obtained by estimating the trend regime parameters from the true trend regime data alone.

Having obtained estimates of $\mathbf{\beta}$ and $\mathbf{\theta}$, we can set $C$ to be on the order of the log-likelihood component in Equation \eqref{eqn: loglikapp} at the estimated values of $\mathbf{\beta}$ and $\mathbf{\theta}$ in order to place approximately  equal weight on model fitting and change point regularization.
While alternative procedures such as using cross-validation to choose $C$ can be used, we find that this simpler strategy performs well empirically by encouraging the inflection point of the transition function to occur in the interval $[\tau_a, \tau_b]$. 
Having now also fixed $C$, we can maximize
\eqref{eqn: loglikapp} with respect to $\mathbf{\alpha}$, $d$, $\mu$, and $\nu^2$. As in T2CD-step, we maximize \eqref{eqn: loglikapp} twice, once for $-0.5 \leq d \leq 0.5$, and a second time for $0.5 \leq d \leq 1.5$, and choose the maximizing set of values of $\mathbf{\alpha}$, $d$, $\mu$, and $\nu^2$ that includes an estimate of $d$ that is further from $0.5$.

\subsection{Exact and Generalized Estimation for Multivariate Data}
\label{sec: multivariate_est}

When there are $p$ replicates of the sequences, the penalized log-likelihood function is a sum of the penalized log-likelihoods of the individual sequences. The only constraint is that the long-memory parameter $d$ is shared across dimensions as described in Section \ref{sec: multivariate}.
Recall that the change locations are allowed to differ across replicates, the number of possible combinations for change locations is $m^p$, where $m$ is the number of time indices in $[\tau_a, \tau_b]$. An exhaustive search for the best combination is often computationally prohibitive. 
For this reason, we use the following two-step procedure. First, we run either T2CD-step or T2CD-sigmoid on each univariate sequence to obtain estimates of $\mathbf{\beta}_j$, $\mathbf{\theta}_j$, $d_j$, $\mu_j$, $\nu^2_j$, and $\tau_j$ for T2CD-step or $\mathbf{\alpha}_j$ for T2CD-sigmoid. Fixing the estimates of $\mathbf{\beta}_j$, $\mathbf{\theta}_j$, and $\tau_j$ for T2CD-step or $\mathbf{\alpha}_j$ for T2CD-sigmoid, we then optimize over $d$, $\mu_j$, $\nu^2_j$, initializing $d$ at the mean univariate estimate across all of the time series $p^{-1}\sum_{j = 1}^p \hat{d}_j$ and $\mu_j$ and $\nu^2_j$ at the univariate estimates.

\section{Simulation Study}
\label{sec:simulations}

We evaluate the performance of T2CD-step and T2CD-sigmoid for estimating $\tau$ and $d$ under several different scenarios, using both univariate and multivariate time series data. We set up the simulations to be similar to the ECIS data described in Section \ref{sec: intro}. 
First, we consider one simple scenario and compare estimates of the change point $\tau$ obtained by T2CD-step and T2CD-sigmoid, in order to examine how generalizing the discrete transition using a smooth transition function affects change point estimation. 
We then consider a broader set of scenarios and compare T2CD-step and T2CD-sigmoid not only to each other but also to several alternative methods.

We simulate univariate time series for comparing T2CD-step and T2CD-sigmoid as follows. Given a fixed change point $\tau$, we simulate trend curves $\boldsymbol f = (f_1, \dots, f_\tau)$ from a mean zero Gaussian process with squared exponential kernel $Cov[f_t, f_s] = 10 \text{exp}\paren{-0.5(s-t)^2}$. We simulate trend regime measurements $y_t = f_t + \eta_t$, where $\eta_t$ are mean zero heteroscedastic measurement errors with standard deviation $\sigma_t = \frac{2-0.1}{\max \set{f_s}_{s=1}^\tau - \min \set{f_s}_{s = 1}^\tau}\left[ f_t - \min \set{f_s}_{s=1}^\tau \right] + 0.1$.
We simulate equilibrium measurements $y_{\tau + 1}, \dots, y_T$ according to a mean-zero FI model with noise variance $\nu = 0.5$ and long memory parameter $d$: $(1 - B)^d y_t = \epsilon_t$, where  $\epsilon_t \sim \text{N}(0, 0.25)$. 
For comparison with the observed ECIS data, we simulate univariate time series of length $T = 400$, which we can think of as 70 hours of data.
For each combination of true change points $\tau$ set to values in the interval $[85,258]$ chosen to correspond to change points at $\{15, 20, \dots, 45\}$ hours and long memory parameter $d \in \left\{-0.25, -0.05, \dots, 1.45\right\}$, we simulate 100 univariate time series.
When applying T2CD-step and T2CD-sigmoid to each simulated univariate time series, we set the candidate range of $\tau$ to $[\tau_a = 10, \tau_b = 50]$, use spline basis of degree 3 with knots at every integer value of $t$ when fitting $\beta$, and knots at every integer multiple of 5 when fitting $\theta$. 
For T2CD-sigmoid, we fix $C=1000$ throughout.
We check the choice of these hyperparameters in Figure \ref{fig: mdck} through residual analysis. Extensive studies on hyperparameter tuning is beyond the scope of this work.

\begin{figure}

\begin{subfigure}[b]{\linewidth}
\includegraphics[width=\linewidth]{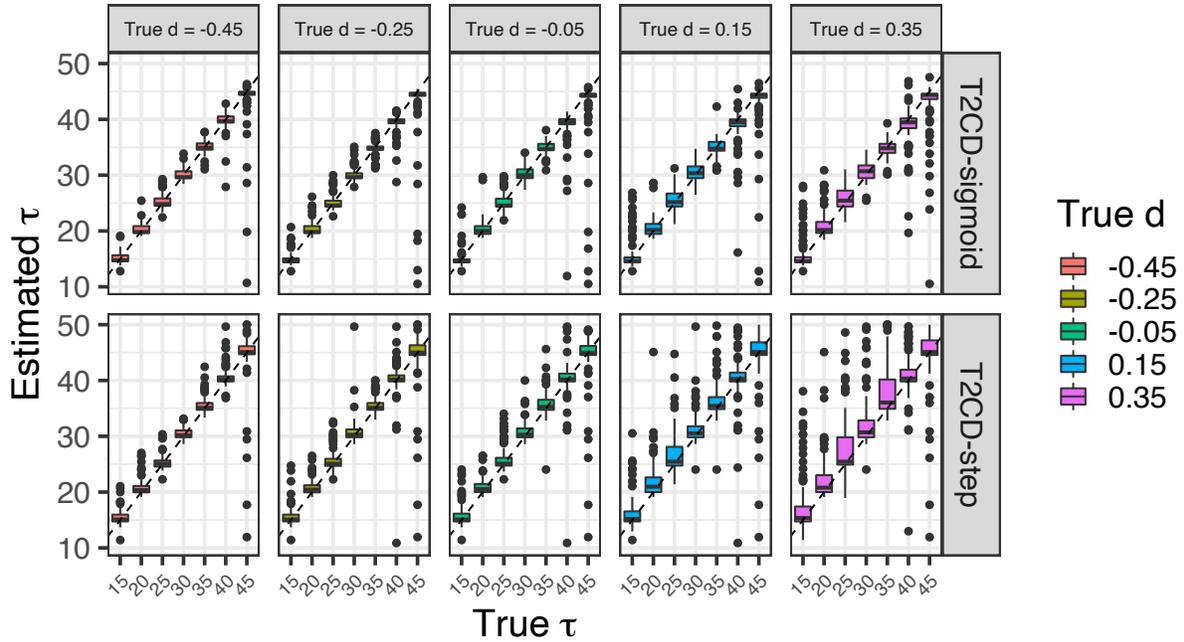}
\caption{Estimates of $\tau$ when the second regime is stationary at $d<0.5$.}
\end{subfigure}

\begin{subfigure}[b]{\linewidth}
\includegraphics[width=\linewidth]{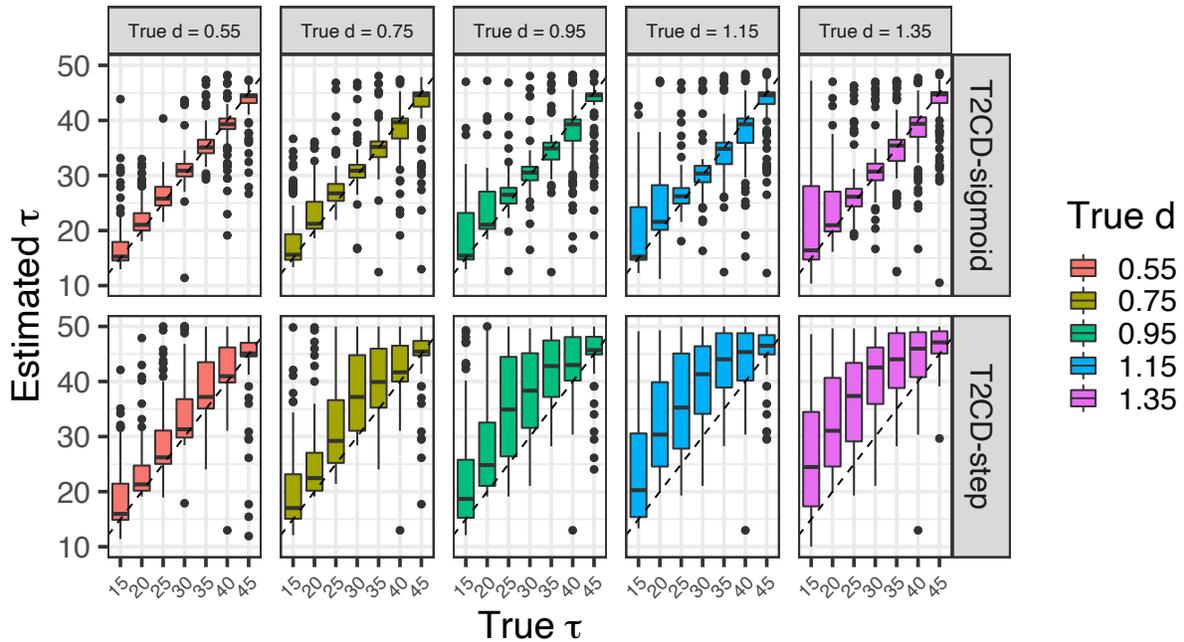}
\caption{Estimates of $\tau$ when the second regime is nonstationary at $d>0.5$. T2CD-step tends to overestimate $\tau$ because nonstationarity can be mistaken for the first regime. The overestimation issue is less severe for T2CD-sigmoid because the smooth transition function accommodates uncertainty about the change point.}
\end{subfigure}

\caption{T2CD estimates of $\tau$ for simulation setup where the first regime is generated via Gaussian process with squared exponential kernel and the second regime generated via FI($d$).}
\label{fig: g2cd_tau}
\end{figure}

The performance of estimates of $\tau$ are shown in Figure \ref{fig: g2cd_tau}. Estimated change points for T2CD-sigmoid are set to the time index when the smooth transition function is equal to 0.5. Both T2CD-step and T2CD-sigmoid estimate the change point $\tau$ well when $d$ is much smaller than $0.5$. 
We hypothesize that when $d$ is close to or larger than $0.5$, the change point is more difficult to recover because the long-range autocorrelations between equilibrium measurements can yield smoothly varying time trends during the equilibrium period, which can be mistaken for a continuation of the trend period.
Surprisingly, when $d$ is closer to or greater than $0.5$ T2CD-step tends to overestimate $\tau$ while T2CD-sigmoid continues to estimate $\tau$ well on average. 

\begin{figure}

\begin{subfigure}[b]{\linewidth}
\includegraphics[width=0.49\linewidth]{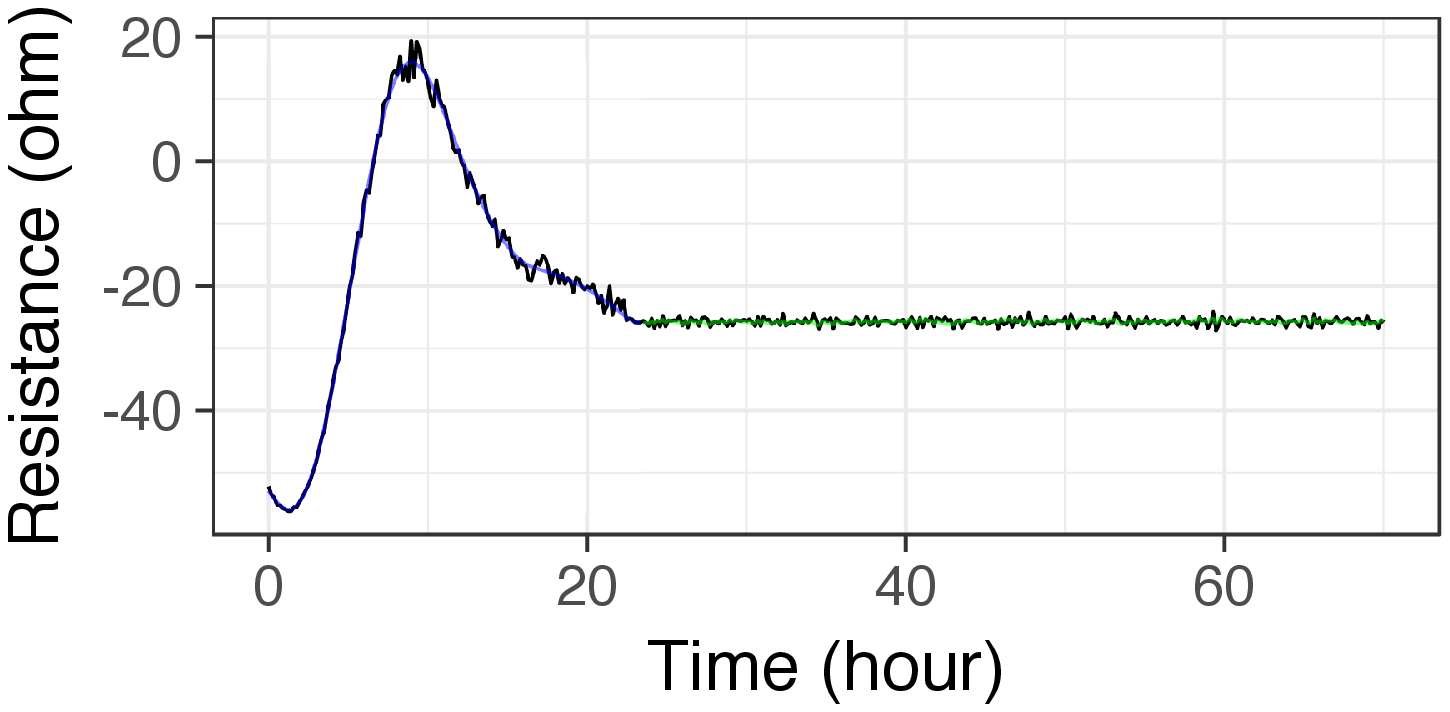}
\includegraphics[width=0.49\linewidth]{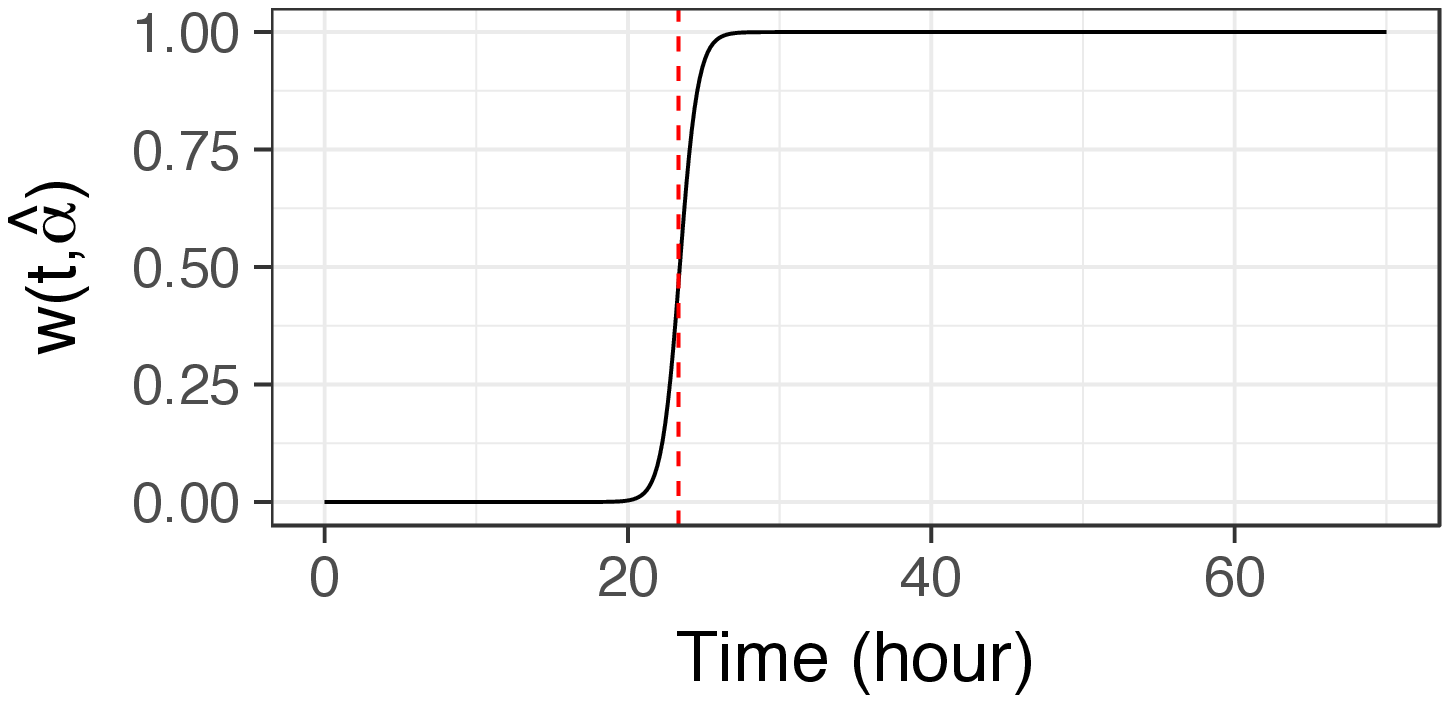}
\caption{Estimates by T2CD-sigmoid for simulation with $d=-0.45$.}
\end{subfigure}

\begin{subfigure}[b]{\linewidth}
\includegraphics[width=0.49\linewidth]{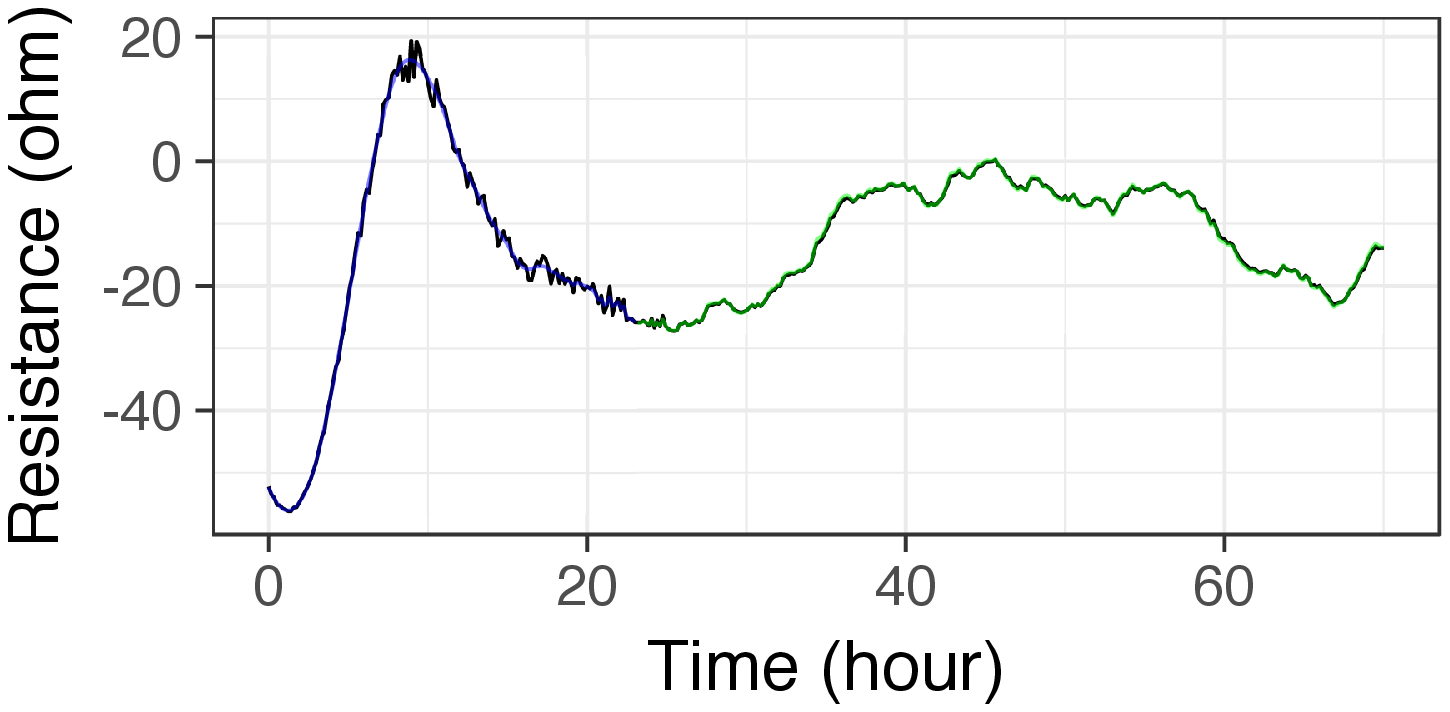}
\includegraphics[width=0.49\linewidth]{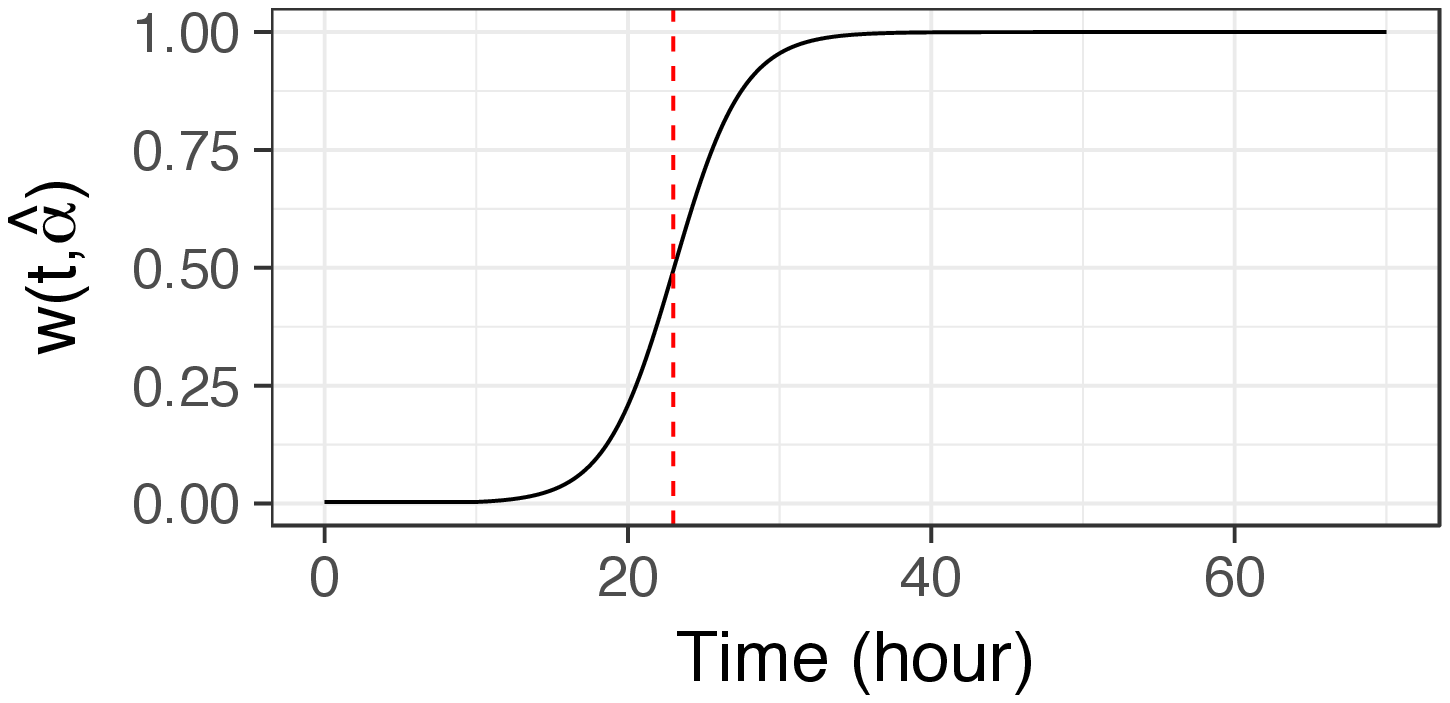}
\caption{Estimates by T2CD-sigmoid for simulation with $d=1.35$. }
\end{subfigure}

\caption{T2CD-sigmoid estimates for simulation setup where the first regime is generated via Gaussian process with squared exponential kernel and the second regime generated via FI($d$). The blue and green overlaid lines are the fit by T2CD-sigmoid for the trend and confluence phase, respectively. The vertical red dashed line marks the time index when the regime transition function is estimated to cross 0.5. The estimated transition is less abrupt for large $d$.}
\label{fig: g2cdfast_wt}
\end{figure}

In order to understand why T2CD-sigmoid provides better change point estimates than T2CD-step when $d$ is close to or greater than $0.5$, we zoom in on a pair of estimated smooth transition functions from simulations with $d = -0.45$ and $d = 1.35$ in Figure \ref{fig: g2cdfast_wt}.
We observe that the estimated transition function is much steeper and more similar to the discrete transition function assumed by T2CD-step when $d = -0.45$.
By allowing a smooth transition function, T2CD-sigmoid can accommodate greater uncertainty about the change point when $d$ is close to or greater than $0.5$.

\begin{figure}
\centering
\includegraphics{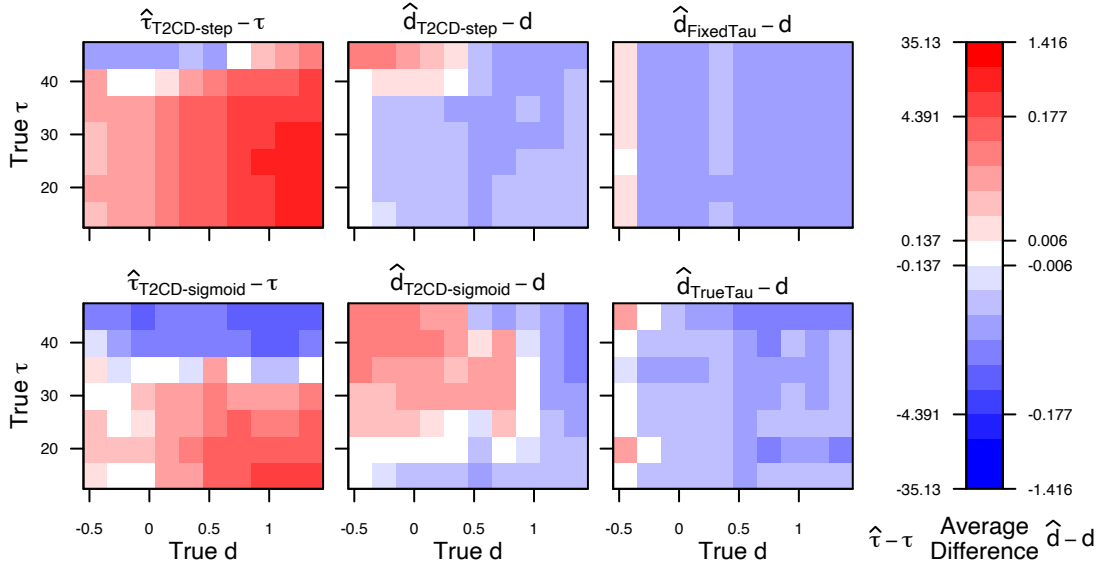}
\caption{Performance of estimates for change location $\tau$ and long-memory parameter $d$ obtained using T2CD-step, T2CD-sigmoid, FixedTau, and TrueTau across 100 simulated series per each combination of $\tau$ and $d$ for each simulation configuration. The first regime is generated via Gaussian process with squared exponential kernel; the second regime generated via FI($d$).}
\label{fig: univgp_eaft}
\end{figure}

Having shown that using T2CD-sigmoid and generalizing the discrete transition function assumed in T2CD-step with a smooth transition function can actually result in improved estimation of the true change point, we examine the relative performance of T2CD-step and T2CD-sigmoid with respect to estimating the long memory parameter $d$ in Figure~\ref{fig: univgp_eaft}. 
For context, we also consider estimation of $d$ using a procedure that fixes the change point $\tau =50$ (FixedTau) and a procedure that fixes the change point $\tau$ at its true value (TrueTau). 
FixedTau sets the bar for estimating $d$ with conservative data usage, whereas TrueTau gives the best $d$ estimate that can be attained if the true change point were known. 

\begin{figure}
\centering
\includegraphics{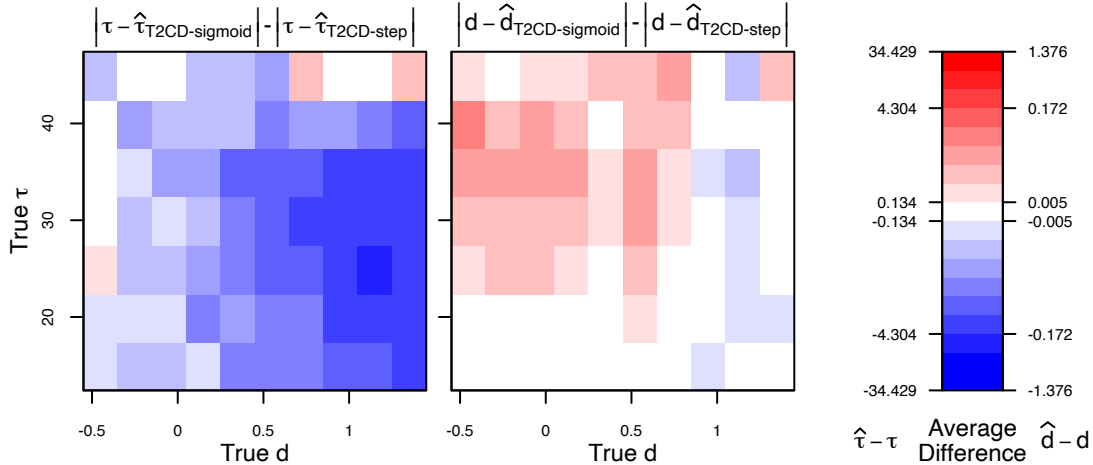}
\caption{Relative performance of estimates for change location $\tau$ and long-memory parameter $d$ obtained using T2CD-step and T2CD-sigmoid across 100 simulated series per each combination of $\tau$ and $d$ for each simulation configuration. The first regime generated via Gaussian process with squared exponential kernel; the second regime generated via FI($d$).}
\label{fig: univgp_ea}
\end{figure}

We see that estimation of the change point $\tau$ and estimation of the long memory parameter $d$ are closely related. When the estimated change point occurs too early, we tend to overestimate the long-memory parameter. When the estimated change point occurs too late, we tend to underestimate the long-memory parameter $d$. 
This pattern is most apparent when T2CD-step is used.
Both T2CD-step and T2CD-sigmoid provide better estimates of $d$ than FixedTau as long as the true change point occurs before 40 hours. We also observe that T2CD-step provides only slightly poorer estimation of $d$ than TrueTau. We further investigate the relative performance of T2CD-step and T2CD-sigmoid in Figure~\ref{fig: univgp_ea}. 

We see that T2CD-step and T2CD-sigmoid provide comparably accurate estimates of the differencing parameter $d$ when the change point occurs early. When the equilibrium process is non-stationary with true long memory parameter $d > 0.5$, the improved estimation of the change point T2CD-sigmoid relative to T2CD-exact also results in improved estimation of the long-memory parameter $d$.

\begin{figure}
\centering
\includegraphics{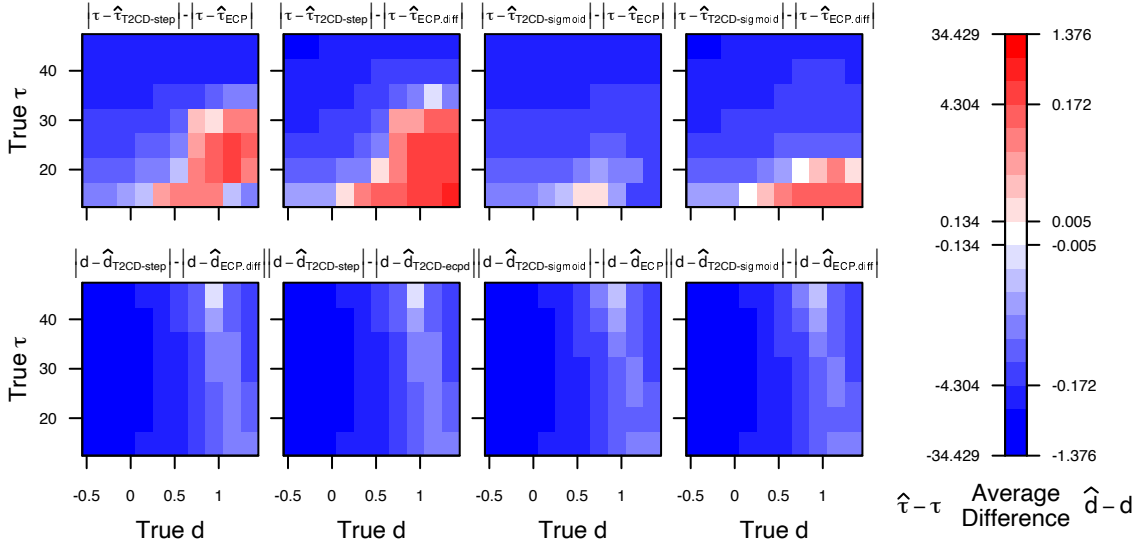}
\caption{Relative performance of estimates for change location $\tau$ and long-memory parameter $d$ obtained using T2CD-step, T2CD-sigmoid, ECP, and ECP.diff across 100 simulated series per each combination of $\tau$ and $d$ for each simulation configuration. The first regime generated via Gaussian process with squared exponential kernel; the second regime generated via FI($d$).}
\label{fig: univgp_ecp}
\end{figure}

Now we compare the performance of T2CD-step and T2CD-sigmoid for estimating the change point $\tau$ and long memory parameter $d$to the performance of two procedures that use the popular E-Divisive algorithm introduced in \cite{matteson2014} to estimate the change point (ECP and ECP.diff).
The popular E-Divisive algorithm is a nonparametric procedure which uses the energy statistics as a distance metric for binary segmentation \citep{matteson2014}. 
E-Divisive can be used to find multiple change points. In our comparison, we use E-Divisive to find a maximum of $3$ change points and use the most significant change point within the candidate range $[\tau_a, \tau_b]$.
We consider two different procedures based on E-Divisive: ECP applies the E-Divisive algorithm to the observed time series $\boldsymbol y$, whereas ECP.diff applies the E-Divisive algorithm to the first difference of the observed time series data. 
Once an estimated change point $\tau$ has been obtained, both ECP and ECP.diff procedures estimate the parameters of the FI model for the equlibrium period using maximum likelihood.
The relative performance of T2CD-step and T2CD-sigmoid compared to ECP and ECP.diff is shown in Figure~\ref{fig: univgp_ecp}.

When compared to alternative methods ECP and ECP.diff, both T2CD-step and T2CD-sigmoid estimate the change point $\tau$ better for all true change points when the equilibrium process is stationary with $d < 0.5$, and for late true change points $\tau > 35$ when the equilibrium process is non-stationary with $d \geq 0.5$.
Careful examination of the change point estimates indicates that ECP and ECP.diff tend to underestimate the change point, which is likely due to the fact that both assume that observations between change points are independently and identically distributed. 
This does not hold for data that we simulated, nor do we expect it to hold for the ECIS data described in Section~\ref{sec: intro}.

Relative performance of the long memory parameter $d$ mirrors the relative performance of the change point $\tau$.
T2CD-step and T2CD-sigmoid tend to perform comparably, with slightly better estimates of the long memory parameter $d$ from T2CD-step when the equilibrium process is stationary with $d < 0.5$ and slightly better estimates of the long memory parameter $d$ from T2CD-sigmoid when the equilibrium process is non-stationary with $d \geq 0.5$.
ECP and ECP.diff produce much poorer estimates of the long memory parameter $d$ than both T2CD-step and T2CD-sigmoid for all true change point and long memory parameter values, which is unsurprising given we observed poorer estimates of the change point $\tau$ from ECP and ECP.diff.

However, the performance advantages of T2CD-step and T2CD-sigmoid do come at a computational price. For the first univariate experiment where the trend regime is generated via Gaussian processes, on average on a 2.7 GHz CPU, ECP and ECP.diff both take 1.20 seconds, T2CD-step takes 196 seconds and T2CD-sigmoid takes 19.2 seconds.
While both of the T2CD methods are slower than the alternatives, T2CD-sigmoid is roughly 10 times faster than T2CD-step on average.
This makes T2CD-sigmoid a competitive option in providing balance between the quality of estimation and computational speed.

\begin{figure}
\centering
\includegraphics{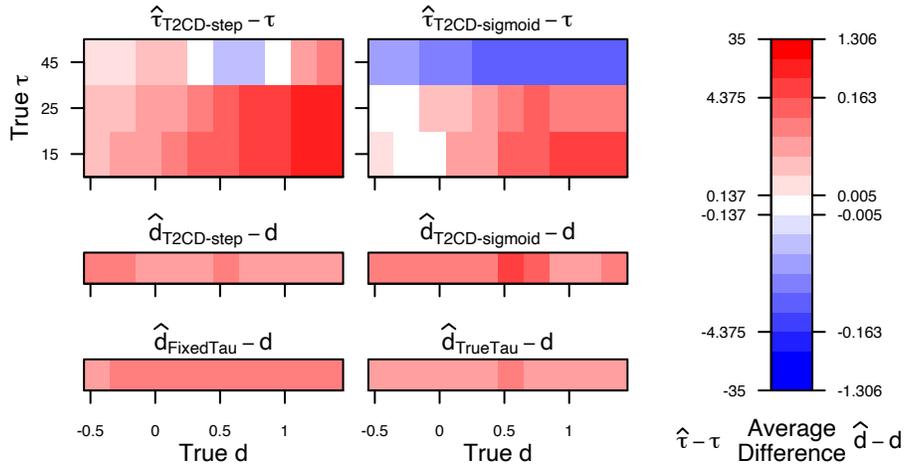}
\caption{Performance of estimates for change location $\tau$ and long-memory parameter $d$ obtained using T2CD-step, T2CD-sigmoid, FixedTau, and TrueTau across 100 simulated series per each combination of $\tau$ and $d$. Multivariate simulations with $p = 3$, with change points at 15, 25, and 45. For each series, the first regime is generated via Gaussian process with squared exponential kernel, and the second regime is generated via FI($d$).}
\label{fig: mv_eaft}
\end{figure}

Next, we consider multivariate time series data made up of $p$ individual time series with unique change points $\tau_1, \dots, \tau_p$ and common long memory parameter $d$. 
We simulate $100$ multivariate time series of length $T = 400$ with $p = 3$ for each value of the long memory parameter $d \in \left\{-0.25, -0.05, \dots, 1.45\right\}$. For each value of $d$, a single simulated multivariate time series is comprised of three individual time series with different change points $\tau_1 = 15$, $\tau_2 = 25$, and $\tau_3 = 45$. As in the univariate simulations, trend curves $\boldsymbol f_j = (f_{j1}, \dots, f_{j\tau_j})$ are simulated from a mean zero Gaussian process with squared exponential kernel $Cov[f_t, f_s] = 10 \text{exp}\paren{-0.5(s-t)^2}$. 
We simulate trend regime measurements $y_{jt} = f_{jt} + \eta_{jt}$, where $\eta_{jt}$ are mean zero heteroscedastic measurement errors with standard deviation $\sigma_{jt} = \frac{2-0.1}{\max \set{f_{js}}_{s=1}^{\tau_j} - \min \set{f_{js}}_{s = 1}^{\tau_j}}\left[ f_{jt} - \min \set{f_{js}}_{s=1}^{\tau_j} \right] + 0.1$.
We simulate equilibrium measurements $y_{\tau_j + 1}, \dots, y_T$ according to a mean-zero FI model with noise variance $\nu = 0.5$ and long memory parameter $d$: $(1 - B)^d y_{jt} = \epsilon_{jt}$, where  $\epsilon_{jt} \sim \text{N}(0, 0.25)$.
Again, we set the candidate range of $\tau$ to $[\tau_a = 10, \tau_b = 50]$, use spline basis of degree 3 with knots at every integer value of $t$ when fitting $\beta$, and knots at every integer multiple of 5 when fitting $\theta$. 
For T2CD-sigmoid, we fix $C=1000$ throughout.
Estimates of the change point $\tau$ and long-memory parameter $d$ are summarized in Figure \ref{fig: mv_eaft}.

The multivariate results shown in Figure~\ref{fig: mv_eaft} mirror the univariate results shown in Figure~\ref{fig: univgp_eaft}. 
Both T2CD-step and T2CD-sigmoid tend to overestimate earlier changepoints and underestimate the latest changepoint.
Also, both T2CD-step or T2CD-sigmoid slightly overestimation the differencing parameter $d$.
The performance of both T2CD-step and T2CD-sigmoid is on par with the conservative and oracle methods FixedTau and TrueTau. T2CD-step and T2CD-sigmoid provide better estimates of the long-memory parameter $d$ than FixedTau as long as the true long-memory parameter is not close to $d = 0.5$, and only slightly worse estimates of the long-memory parameter $d$ than TrueTau.

\begin{figure}
\centering
\includegraphics{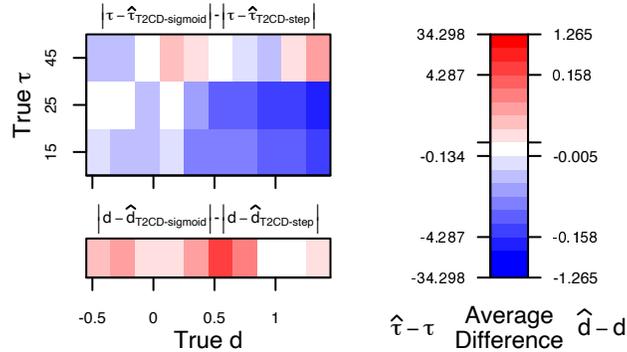}
\caption{Relative performance of estimates for change location $\tau$ and long-memory parameter $d$ obtained using T2CD-step and T2CD-sigmoid across 100 simulated series per each combination of $\tau$ and $d$. Multivariate simulations with $p = 3$, with change points at 15, 25, and 45. For each series, the first regime is generated via Gaussian process with squared exponential kernel, and the second regime is generated via FI($d$).}
\label{fig: mv_ea}
\end{figure}

Figure~\ref{fig: mv_ea} zooms in on the relative performance of T2CD-step and T2CD-sigmoid. 
T2CD-sigmoid tends to provide better estimation of the change points $\tau_1$, $\tau_2$, and $\tau_3$. Better estimation of the differencing parameter $d$ is provided by T2CD-step when the equilibrium process is more stationary. 

\begin{figure}
\centering
\includegraphics{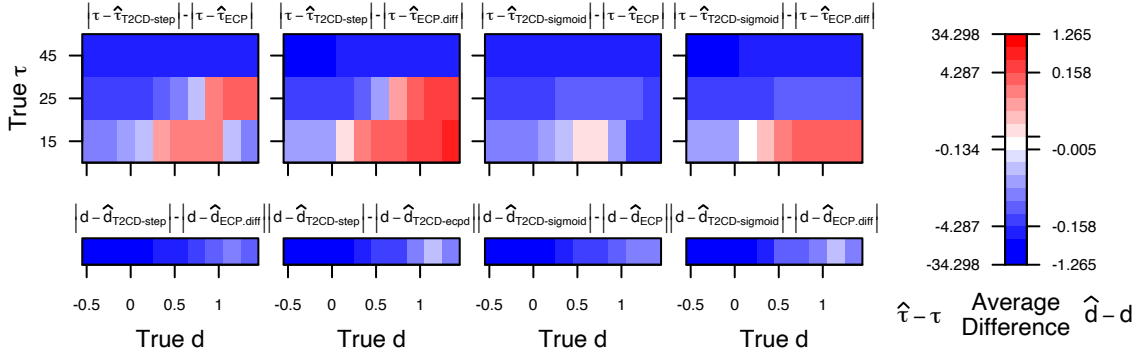}
\caption{Relative performance of estimates for change location $\tau$ and long-memory parameter $d$ obtained using T2CD-step, T2CD-sigmoid, ECP, and ECP.diff across 100 simulated series per each combination of $\tau$ and $d$. Multivariate simulations with $p = 3$, with change points at 15, 25, and 45. For each series, the first regime is generated via Gaussian process with squared exponential kernel, and the second regime is generated via FI($d$).}
\label{fig: mv_vs_ecp}
\end{figure}

Figure~\ref{fig: mv_vs_ecp} examines the relative performance of T2CD-step and T2CD-sigmoid compared to ECP and ECP.diff. T2CD-step and T2CD-sigmoid provide better estimates of the change points $\tau_1$, $\tau_2$, and $\tau_3$ compared to ECP and ECP.diff, as long as the equilibrium process is stationary or the change point occurs late.
Regarding estimation of the long-memory parameter $d$, we observe consistently better performance of T2CD-step and T2CD-sigmoid estimates relative to ECP and ECP.diff estimates.

\section{Application to ECIS Data}
\label{sec:ecis}

Now we apply the T2CD-step and T2CD-sigmoid to the MDCK and BSC cell data described in Section~\ref{sec: intro}.
ECIS resistance measurements were obtained at several frequencies, however we focus on resistance measured at the frequency of 500 hertz. We also exclude wells that are mechanically disrupted to create a ``wound-healing'' assay and a single well containing MDCK cells that displayed evidence of instrument failure.
In order to assess whether or not cell culture preparation affects our ability to identify cells contaminated with mycoplasma,
we analyze data from BSA and gel wells separately.

For model fitting, we use spline basis of degree 3 with knots at every integer value of $t$ when fitting $\beta$, and knots at every integer multiple of 5 when fitting $\theta$.
As in Section~\ref{sec:simulations}, we set $C=1000$ when implementing T2CD-sigmoid.
Based on visual inspection of the MDCK and BSC data, we set the candidate range of $\tau$ is set to $[\tau_a = 10, \tau_b = 50]$ for MDCK cells and $[\tau_a = 5, \tau_b = 45]$ for BSC cells.

\begin{figure}

\begin{subfigure}[b]{\linewidth}
\centering
\includegraphics[width=0.45\linewidth]{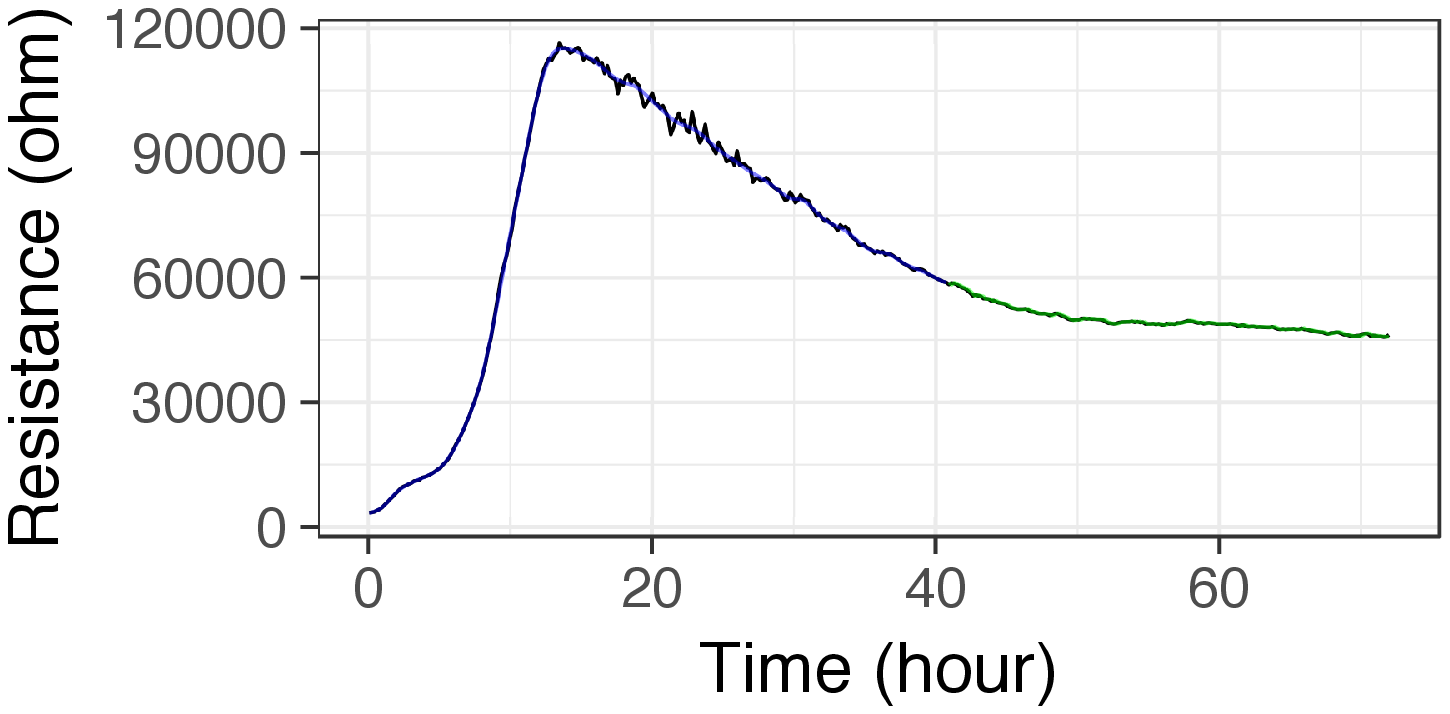}
\caption{Time series of resistance measurements recorded at 500 hertz. The blue and green overlaid lines are the fit by T2CD-step for the trend and confluence phase, respectively.}

\end{subfigure}
\begin{subfigure}[b]{\linewidth}
\includegraphics[width=0.32\linewidth]{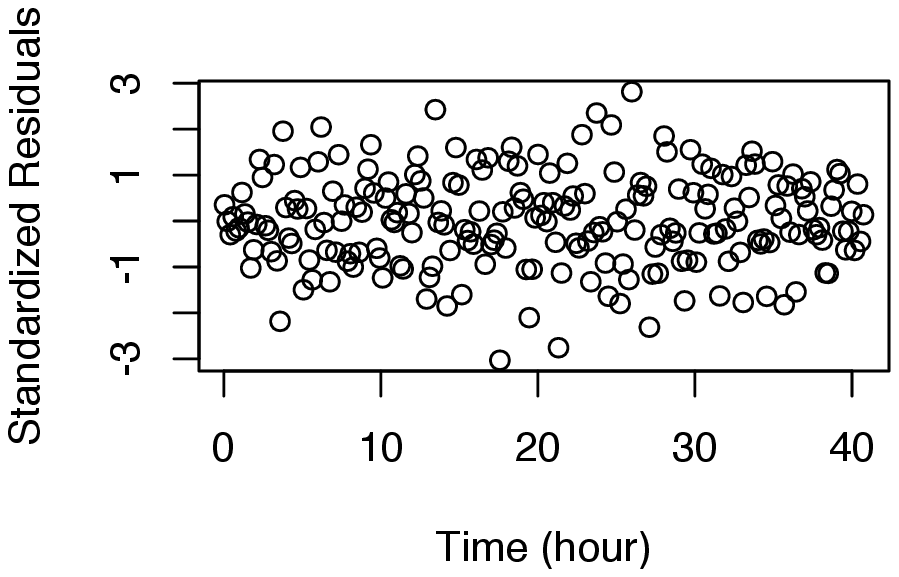}
\includegraphics[width=0.32\linewidth]{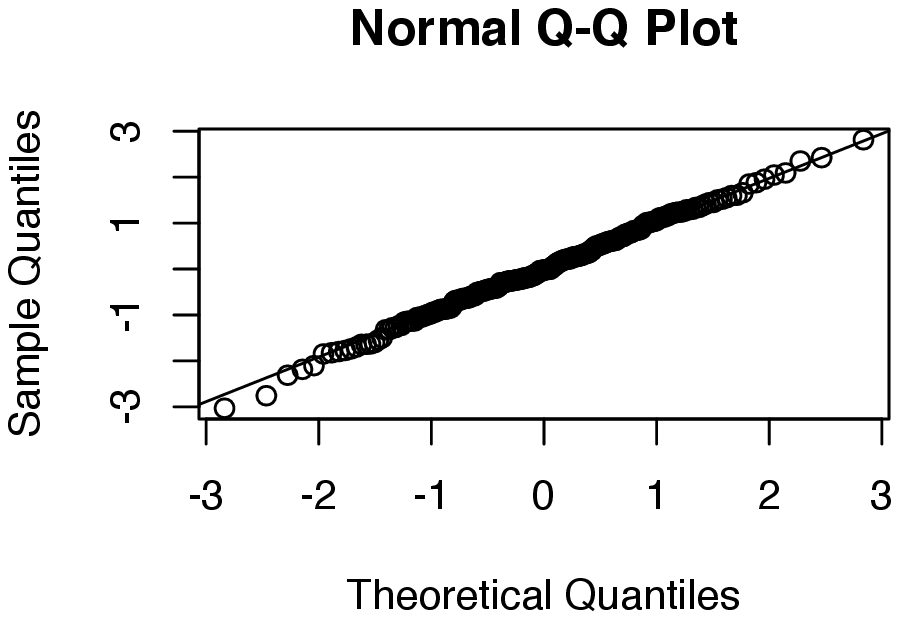}
\includegraphics[width=0.32\linewidth]{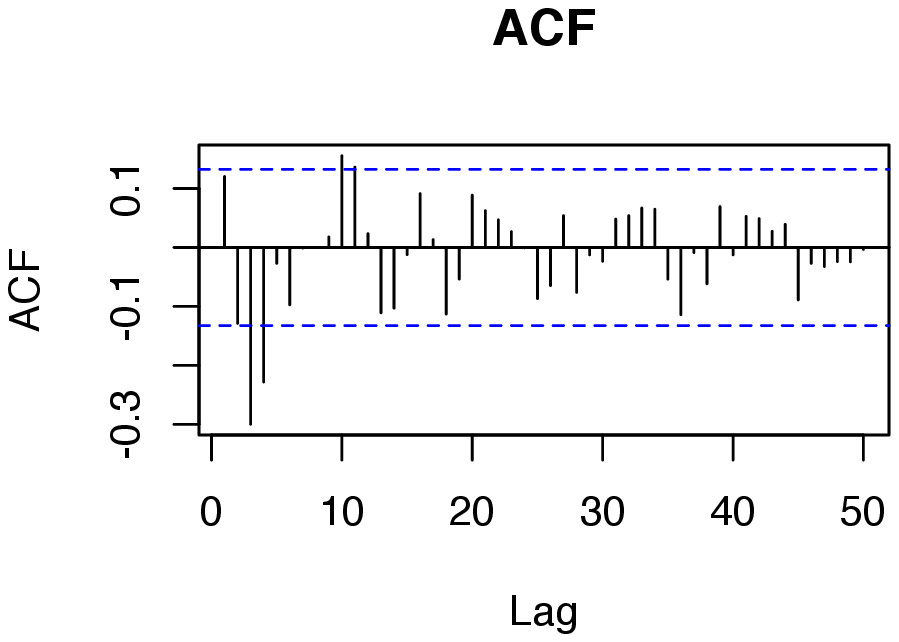}
\caption{Standardized residuals for the trend phase.}
\end{subfigure}

\begin{subfigure}[b]{\linewidth}
\includegraphics[width=0.32\linewidth]{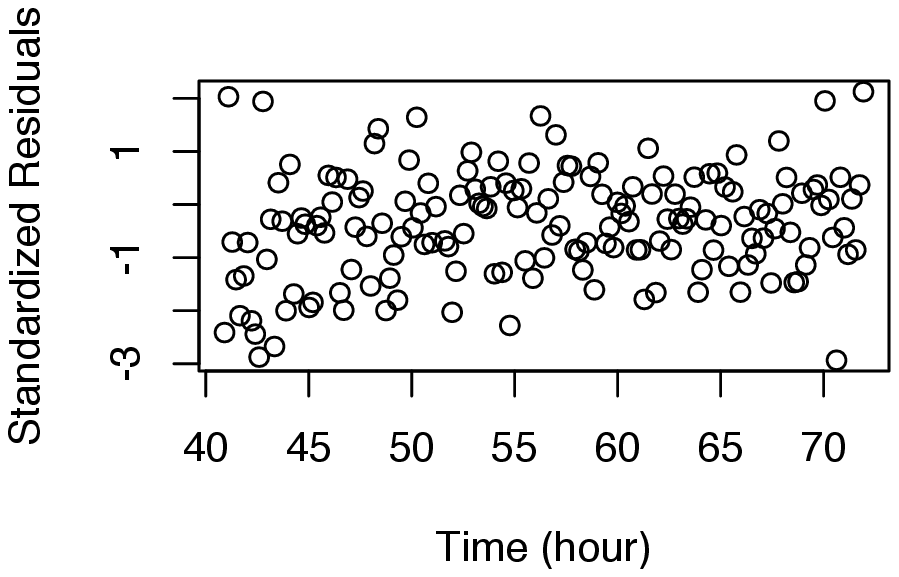}
\includegraphics[width=0.32\linewidth]{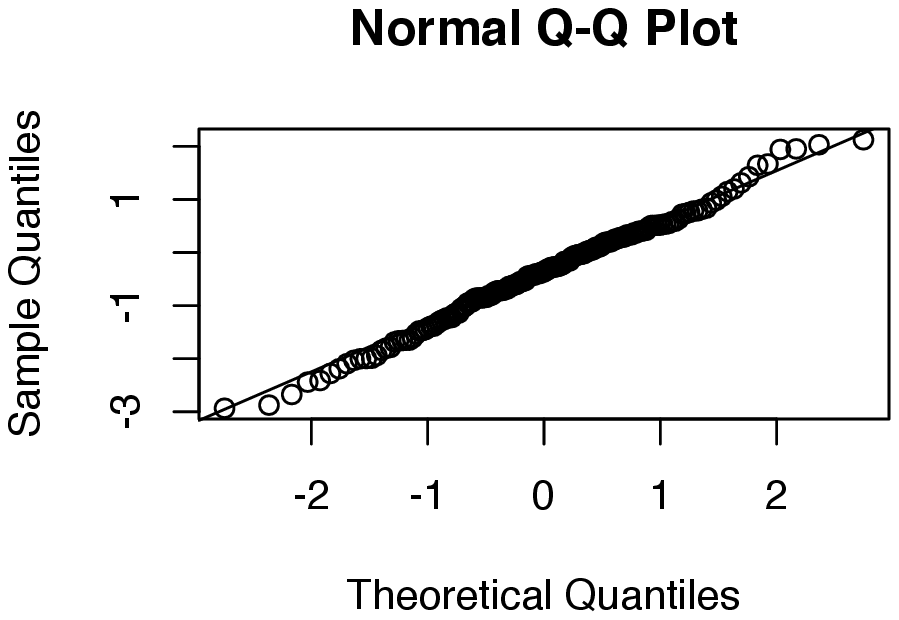}
\includegraphics[width=0.32\linewidth]{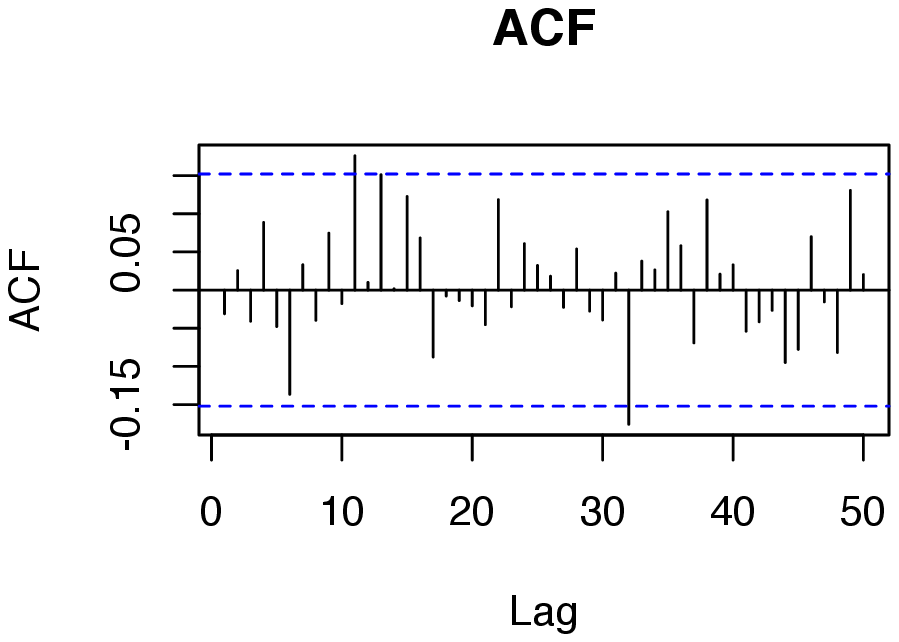}
\caption{Standardized residuals for the confluence phase.}
\end{subfigure}

\caption{MDCK cell; infected and cultivated in gel.}
\label{fig: mdck}
\end{figure}

To check the choice of hyperparameters, we plot in Figure \ref{fig: mdck} the time series for a MDCK sample, as well as standardized residuals from model fitting with T2CD-step. The residuals from the first regime are scaled by $\sigma_t$ estimated. The plots show that our choice of model parameters give reasonable fits to both regimes of the data.

\subsection{MDCK cell line}

From Figure \ref{fig: norinf}, we see that the resistance measurements for MDCK cells  tend to peak before slightly decreasing and stablizing. The start of confluence or equilibrium is hypothesized to be at or slightly after the peak, and as a result is visually distinct.  

\begin{figure}
    \centering
    
    \begin{subfigure}[b]{\linewidth}
    \includegraphics{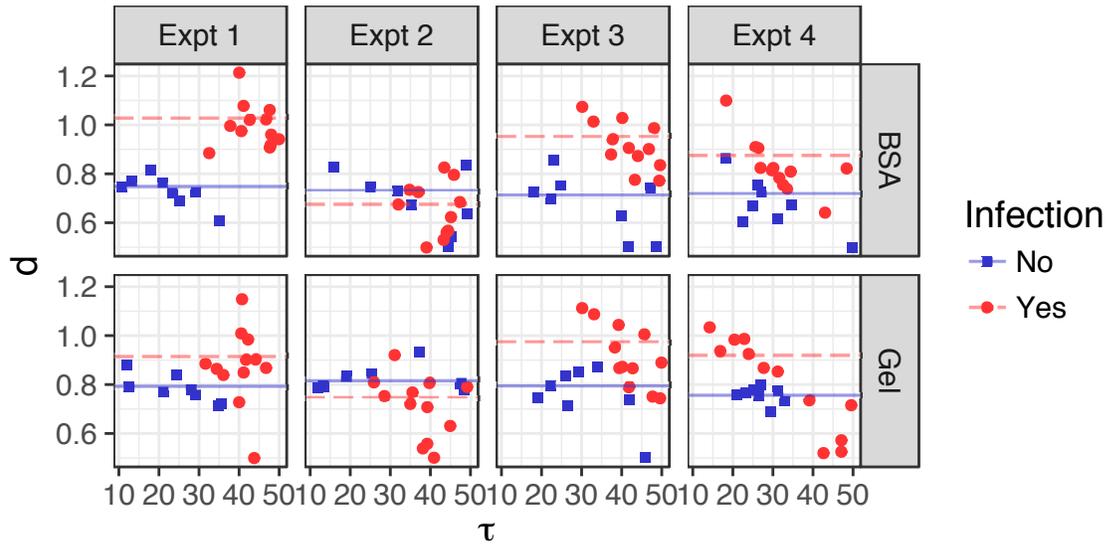}
    \caption{MDCK}
    \label{fig: mdck_exact}
    \end{subfigure}
    
    \begin{subfigure}[b]{\linewidth}
    \includegraphics{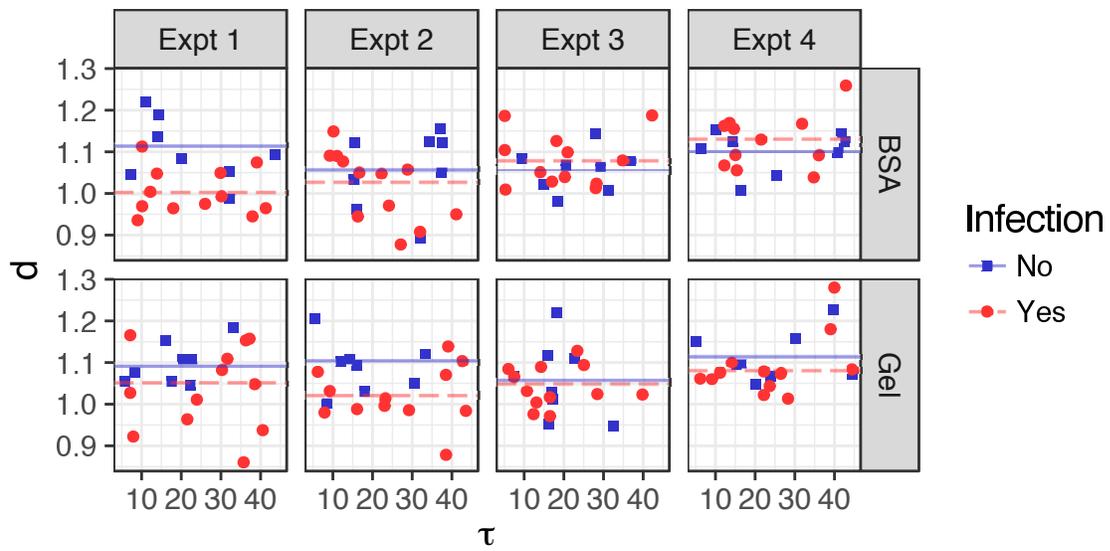}
    \caption{BSC}
    \label{fig: bsc_exact}
    \end{subfigure}
    
    \caption{T2CD-step estimates for $\tau$ and $d$. Points are estimates from the univariate version of the method, and horizontal lines mark estimates from the multivariate version.}
\end{figure}

Figure \ref{fig: mdck_exact} plots estimates of the change points $\tau$ and long memory parameters $d$ estimated by applying T2CD-step to each well as a univariate time series, and by applying T2CD-step to all replicate wells within the same experiment as a multivariate time series.
The estimated change points are scattered within the candidate range of $[10, 50]$, signifying varied initial conditions even in the same batch.
We observe clear evidence of long-range dependence at confluence, with all estimates of the long memory parameter above $0.5$.
Experiments 1, 3, and 4 suggest that MDCK cells that are contaminated by mycoplasma tend to show longer memory than MDCK cells that are uncontaminated.
Experiment 2 suggests the opposite, but this may be a consequence of batch effects. 
Web Appendix C contains more a detailed  review of estimates of the change point $\tau$ and the long memory parameter $d$ estimated by T2CD-step and T2CD-sigmoid across experiments, serum types and infection status.

\subsection{BSC Cells}
\label{sec: bsc}

From Figure \ref{fig: norinf}, we see that the resistance measurements for the BSC cell line tend to increase sharply before plateauing. As compared to the MDCK cell line, the end of the BSC trend phase is less visually distinct. This makes change point detection and subsequent estimation of the long-memory parameter more difficult.

Figure \ref{fig: bsc_exact} plots estimates of the change points $\tau$ and long memory parameters $d$ estimated by applying T2CD-step to each well as a univariate time series, and by applying T2CD-step to all replicate wells within the same experiment as a multivariate time series.
We observe evidence of long memory regardless of contamination status, with most univariate and multivariate estimates of the long memory parameter $d$ exceeding one.
We do not observe distinct separation between the contaminated and uncontaminated cells. 
However, we do observe some evidence that contaminated BSC cells tend to have shorter memory, corresponding to lower estimates of $d$, than uncontaminated cells in Experiments 1 and 2. 
See Web Appendix C for a more detailed review of estimates of the change point $\tau$ and the long memory parameter $d$ estimated by T2CD-step and T2CD-sigmoid across experiments, serum types and infection status.

\subsection{Mycoplasma Contamination Classification}

To demonstrate the quality and utility of our change point $\tau$ and long memory parameter $d$ estimates, we incorporate the estimates as features in a downstream task of classifying cells by their mycoplasma contamination status. 
We build on the linear discriminant analysis (LDA) and quadratic discriminant analysis (QDA) classifiers built to to classify cell lines using ECIS measurements in \cite{gelsinger2017}. 

Let $c$ indicate the possible classes of observations, which in this application corresponds to whether or not a well contains cells contaminated by mycoplasma. $\boldsymbol z$ a vector of features and $\bar{\boldsymbol z}_c$ be the average feature vector across all observations in  class $c$, LDA and QDA class discriminant scores can both be written as special cases of
\begin{align}
    \delta_c\paren{\boldsymbol z} &= \paren{\boldsymbol z-\widebar{\boldsymbol z}_c}^T \widehat{\Sigma}_c^{-1}(\rho)\paren{\boldsymbol z-\widebar{\boldsymbol z}_c} + \log |\widehat{\Sigma}_c(\rho)|\\
    \widehat{\Sigma}_c(\rho) &= (1-\rho)\widehat{\Sigma}_c + \rho\widehat{\Sigma} \nonumber
\end{align}
LDA is obtained by setting $\rho = 1$ and QDA is obtained by setting $\rho = 0$.

For each cell line we train four LDA and QDA classifiers, training each classifier on data from three experiments and computing classification accuracy on data from the remaining experiment.
The average classification accuracy across all four classifiers is provided in Table \ref{tab: classify}, along with the correspond standard deviations. 
We compare classifiers trained using the original features described in \cite{gelsinger2017} to classifiers trained using the best feature from among the original features described in \cite{gelsinger2017} as well as estimates of the change point $\tau$ and long memory parameter $d$, obtained by applying either T2CD-step or T2CD-sigmoid to data from each well as a univariate time series.
A more detailed description of how we constructed the original features described in \cite{gelsinger2017} for our ECIS measurements is given in the Web Appendix D.

\begin{table}[]
\centering
\begin{tabular}{llllll}
\hline
\multirow{2}{*}{Cell line} & \multirow{2}{*}{Features} & \multicolumn{2}{l}{LDA}               & \multicolumn{2}{l}{QDA} \\ \cline{3-6} 
                           &                           & Mean  & SD                           & Mean      & SD         \\ \hline
\multirow{3}{*}{MDCK}      & Original                  & 0.743 & 0.272 & 0.580 & 0.237       \\ \cline{2-6}                        
                           & T2CD-step                & 0.880 & 0.091 & 0.862 & 0.136       \\ \cline{2-6}
                           & T2CD-sigmoid                 & 0.962 & 0.033 & 0.975 & 0.020       \\ \hline
\multirow{3}{*}{BSC}       & Original                  & 0.563 & 0.060 & 0.588 & 0.072       \\ \cline{2-6} 
                           & T2CD-step                & 0.650 & 0.098 & 0.630 & 0.113       \\ \cline{2-6} 
                           & T2CD-sigmoid                 & 0.675 & 0.108 & 0.644 & 0.085      \\ \hline
\end{tabular}
\caption{Classification accuracy for infection status. Average is taken by taking each of the 4 experiments as the test set, and the other 3 as training set. Parameters $\tau$ and $d$ estimated by T2CD increased classification accuracy for both MDCK and BSC cell line.}
\label{tab: classify}
\end{table}

From Table \ref{tab: classify}, it is evident that the $\tau$ and $d$ estimates from T2CD-step and T2CD-sigmoid are useful features that increase classification accuracy for both cell lines. 
For MDCK cells, LDA using the original features has a mean classification accuracy of 0.743. Including T2CD-step or T2CD-sigmoid features improved the mean classification accuracy by 18.4\% and 29.4\%, respectively. For BSC cells, QDA using the original features has a mean classification accuracy of 0.588. Including T2CD-step and T2CD-sigmoid features improved the mean classification accuracy by 7.1\% and 9.5\%, respectively.
The smaller improvements in classification accuracy for BSC cells are likely a consequence of less visually obvious change points in the ECIS measurements for BSC cells, as noted in Section \ref{sec: bsc}.

\section{Conclusion}
\label{sec: conclusion}

In this paper, we propose a model called T2CD for estimating a change point between a smooth, nonlinear trend period and a long-memory equilibrium period and for quantifying features of the trend and equilibrium periods. We provide exact and generalized estimation strategies, T2CD-step and T2CD-sigmoid. Via simulations, we show that T2CD-step outperforms a two step comparison method based on the popular E-Divisive algorithm for change point detection when the equilibrium period can be characterized by a long memory time series model. Compared to E-Divisive, T2CD-step tends to produce better estimates of the change points and long memory parameters. We also show that T2CD-sigmoid offers computational efficiency gains over T2CD-step with minimal reductions and even occasional improvements in performance. 

Practical usage on the MDCK and BSC cell lines shows that T2CD recovers meaningful estimates of change points and long-memory parameters during confluence confluence phase. Importantly, using T2CD reduces the amount of human supervision needed to manually identify change points, ensures that the change points are identified using the same logic, and makes full use of the available data. Furthermore, we show that estimates of the change points and long memory parameters improve classification performance downstream.

\bibliographystyle{apalike}
\bibliography{biblio}

\appendix

\begin{appendices}

\section{T2CD}

\subsection{Feasible Generalized Least Squares}

Heteroscedasticity is addressed through Feasible Generalized Least Squares (FGLS). 
We adopt the following iterative procedure to fit the trend and noise components:
\begin{enumerate}
    \item Estimate the parameters $\beta$ in $f\paren{t; \beta}$ assuming homogeneous noise by penalized least squares;
    \item Estimate the noise standard deviation $\set{\sigma_t}_{t=1}^T$ given $\hat{\beta}_{LS}$ from step 1;
    \item Re-estimate $\beta$ given $\set{\hat{\sigma}_t}_{t=1}^n$ from step 2 through FGLS.
\end{enumerate}

Let $M$ denote the penalty on $\beta$. Then an application of FGLS \citep{yamano2009} to the penalized least squares problem is
\begin{equation*}
    \hat{\beta}_{FGLS} = \paren{X'\widehat{W}^{-1}X + \lambda M}^{-1}X'\widehat{W}^{-1}y
\end{equation*}
where $\widehat{W}$ is an estimate of $W$. 
For the initial estimate of $\beta$, we set $W=I$ under the assumption of homogeneous noise to obtain $\hat{\beta}_{LS}$.
We then fit another spline to the log squared residuals $\log\left(y_t - X_{t\cdot}\hat{\beta}_{LS}\right)^2$, and finally taking the exponential of the spline fit to obtain $\hat{\sigma}_t^2$.
For the re-estimate of $\beta$, we set $\widehat{W}_{t,t} = \hat{\sigma}_t^2$.
Step 2 and 3 can be iterated till convergence.

\subsection{Regime 1 parameter estimation for T2CD-sigmoid}

\begin{figure}
\begin{subfigure}[b]{\linewidth}
\centering
\includegraphics[width=0.8\linewidth]{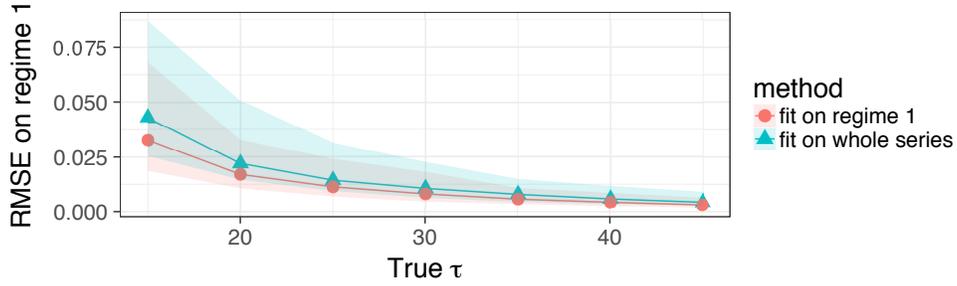}
\caption{The first regime generated via degree-5 polynomials.}
\end{subfigure}

\begin{subfigure}[b]{\linewidth}
\centering
\includegraphics[width=0.8\linewidth]{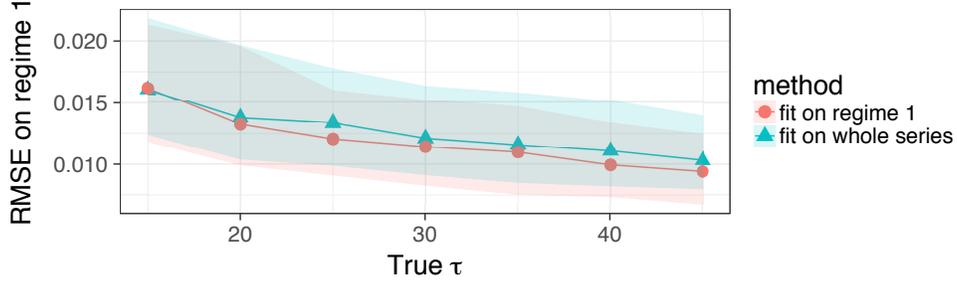}
\caption{The first regime generated via Gaussian process with squared exponential kernel.}
\end{subfigure}

\caption{RMSE on trend estimation by fitting the first regime with the entire series versus the true first regime across 2000 simulated series (100 simulated series for each of 20 $d$'s) per $\tau$. Medians are plotted and error bars indicate the upper and lower quartiles.}
\label{fig: fit1}
\end{figure}

First regime parameters are estimated with the same penalized B-splines procedures as in T2CD-step applied on the entire time series. 
The spline bases are flexible to fit local trends and we show through simulations that fitting on the entire series is almost as good as fitting on the true first regime in Figure \ref{fig: fit1}.

\section{Simulation Study}

\subsection{Polynomial Model for First Regime}

\begin{figure}
\includegraphics{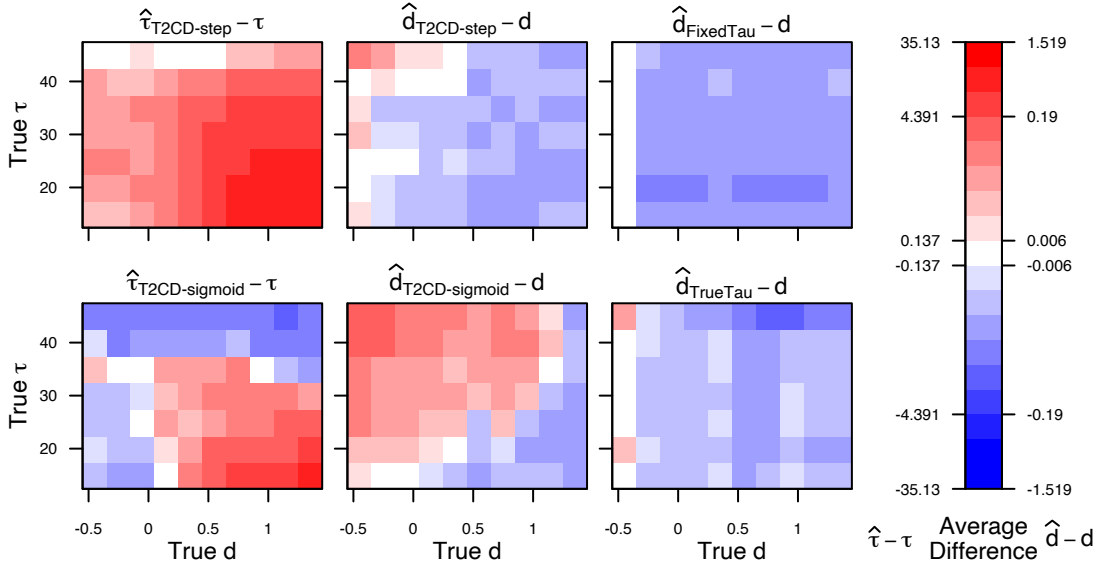}
\caption{Performance of estimates for change location $\tau$ and long-memory parameter $d$ across 100 simulated series per each combination of $\tau$ and $d$. The first regime generated a degree five polynomial; the second regime generated via FI($d$).}
\label{fig: univpoly_eaft}
\end{figure}

\begin{figure}
\includegraphics{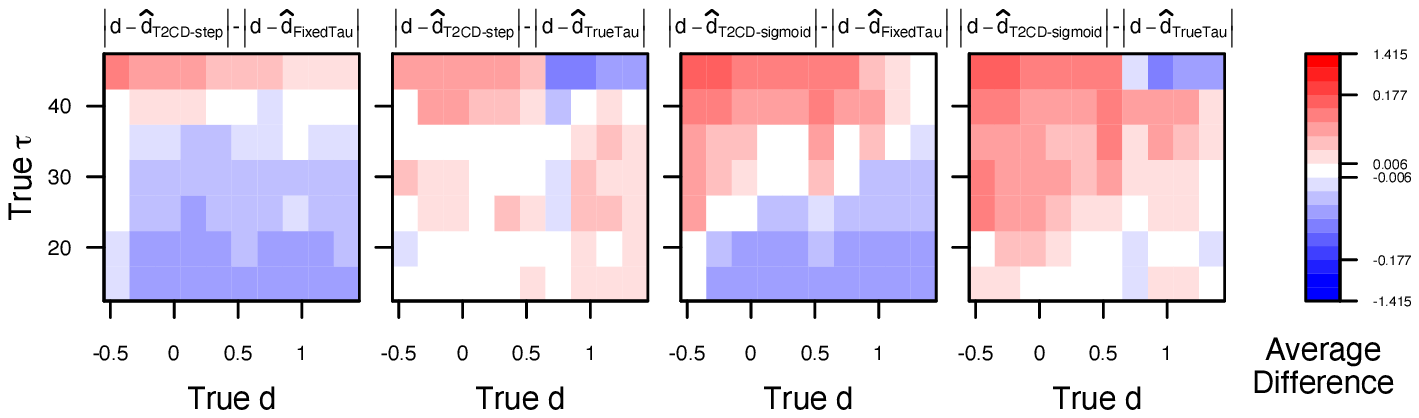}
\caption{Performance of estimates for change location $\tau$ and long-memory parameter $d$ across 100 simulated series per each combination of $\tau$ and $d$. The first regime generated a degree five polynomial; the second regime generated via FI($d$).}
\label{fig: univpoly_ft}
\end{figure}

\begin{figure}
\includegraphics{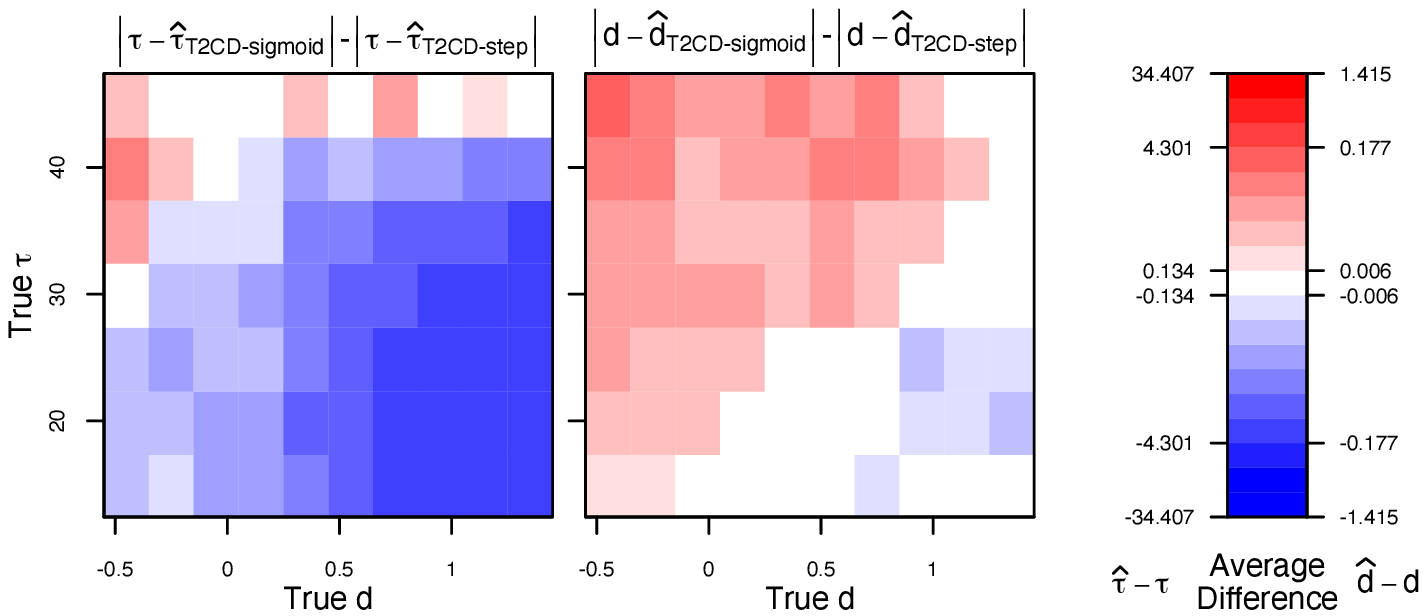}
\caption{Performance of estimates for change location $\tau$ and long-memory parameter $d$ across 100 simulated series per each combination of $\tau$ and $d$. The first regime generated a degree five polynomial; the second regime generated via FI($d$).}
\label{fig: univpoly_ea}
\end{figure}

As a proof-of-concept, we conducted a  simulation study in which we simulate trend regime measurements from a degree-five polynomial model and equilibrium measurements from a FI$\left(d\right)$ model.
The coefficients of the degree-five polynomials are produced by drawing randomly from a normal distribution $N(0, 0.1^2)$ and further shrinking the output for coefficients corresponding to high degrees. For degree $j$ where $j\geq 3$, the output is shrunk by a factor of $0.1^j$. This ensures an expressive series that is not dominated by higher-order terms, since the higher-order terms can result in the first regime being scaled disproportionately to the second regime.
Figures~\ref{fig: univpoly_eaft}-\ref{fig: univpoly_ea} shows the corresponding estimates of $\tau$ and $d$ obtained using T2CD-step, T2CD-sigmoid, ECP, ECP.diff, FixedTau, and TrueTau.

\begin{figure}
\includegraphics{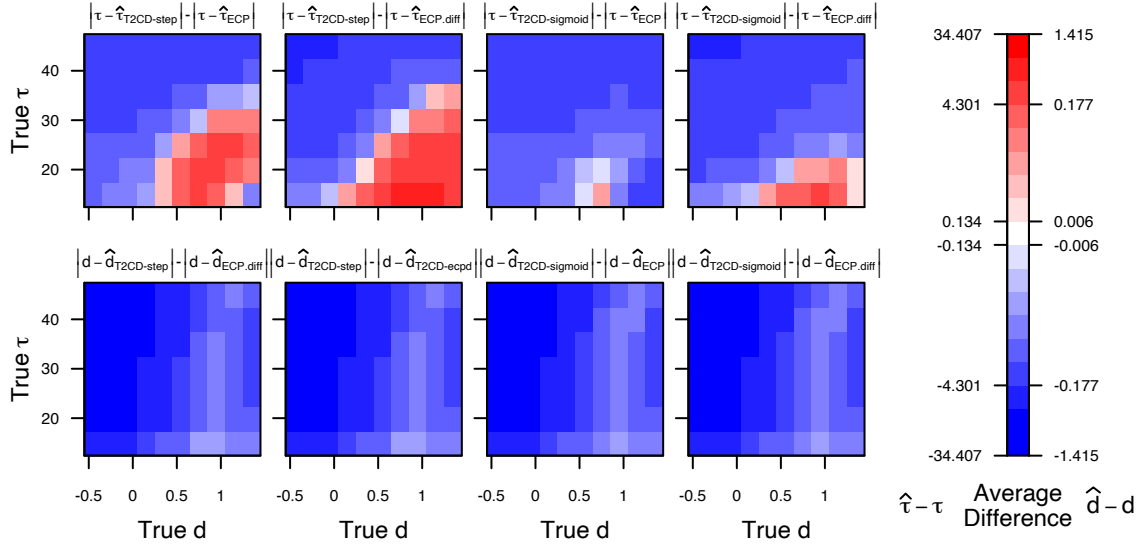}
\caption{Performance of estimates for change location $\tau$ and long-memory parameter $d$ across 100 simulated series per each combination of $\tau$ and $d$. The first regime generated a degree five polynomial; the second regime generated via FI($d$).}
\label{fig: univpoly_ecp}
\end{figure}

\subsection{ARFIMA Model for Second Regime}

\begin{figure}
\includegraphics{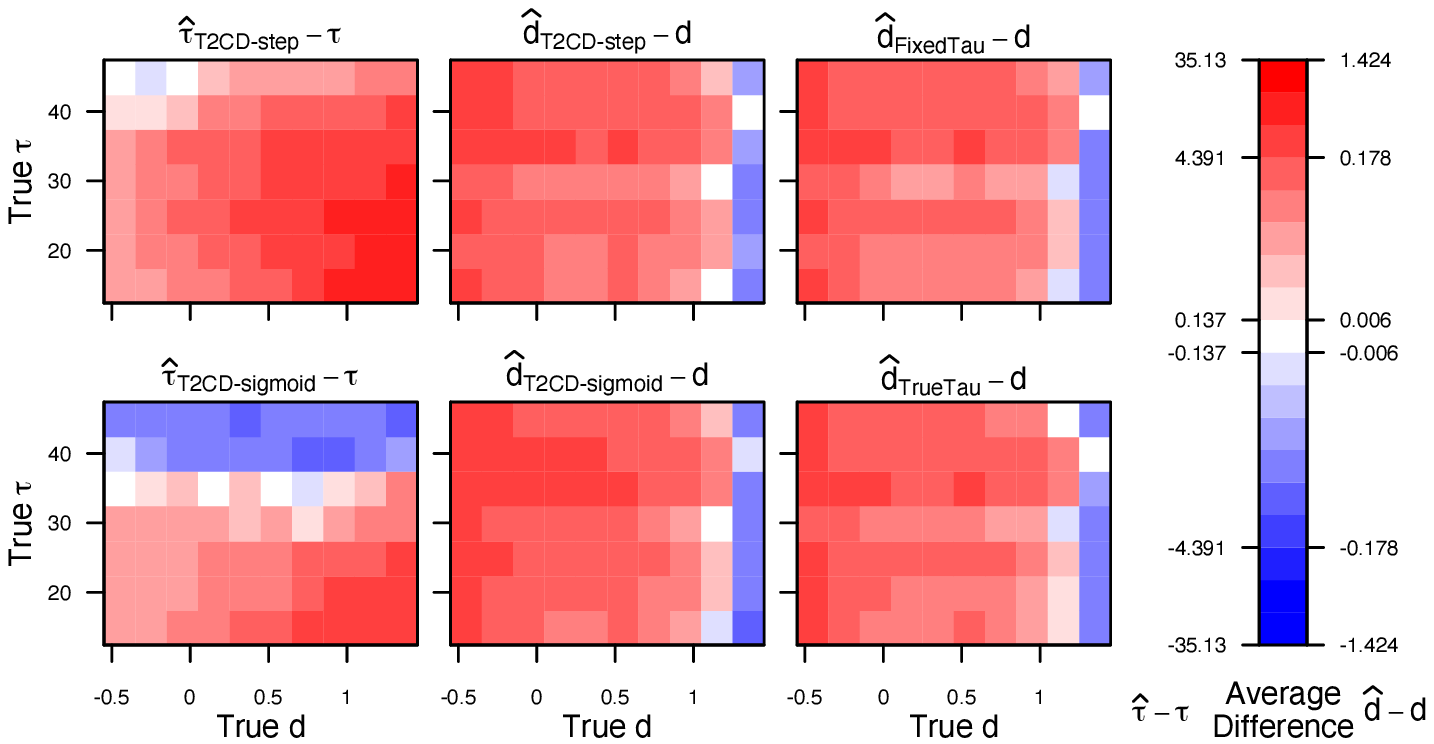}
\caption{Performance of estimates for change location $\tau$ and long-memory parameter $d$ across 100 simulated series per each combination of $\tau$ and $d$. The first regime generated via Gaussian process with squared exponential kernel; the second regime generated via ARFIMA(1,$d$,1).}
\label{fig: univgp_arfima_eaft}
\end{figure}

\begin{figure}
\includegraphics{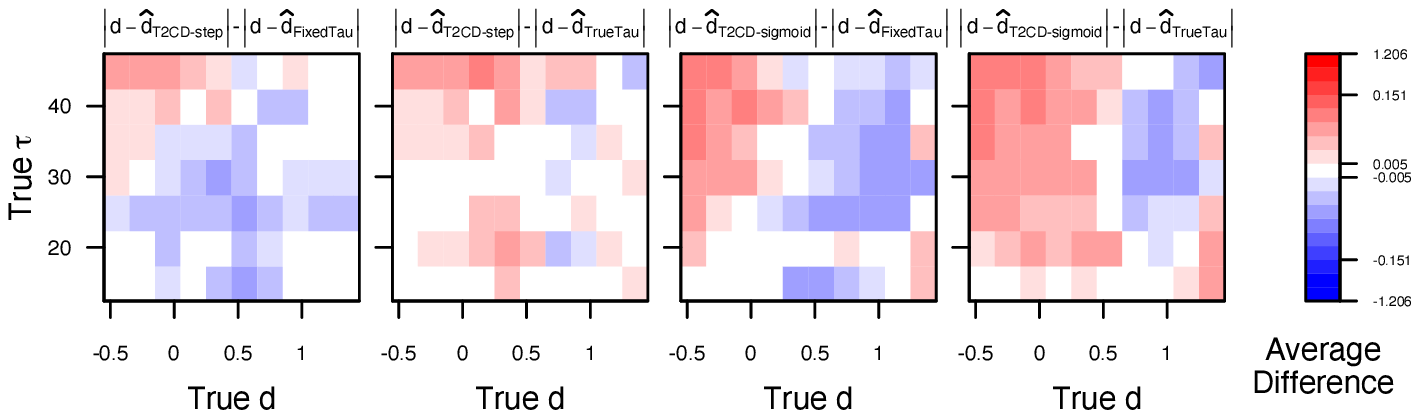}
\caption{Performance of estimates for change location $\tau$ and long-memory parameter $d$ across 100 simulated series per each combination of $\tau$ and $d$. The first regime generated via Gaussian process with squared exponential kernel; the second regime generated via ARFIMA(1,$d$,1).}
\label{fig: univgp_arfima_ft}
\end{figure}

\begin{figure}
\includegraphics{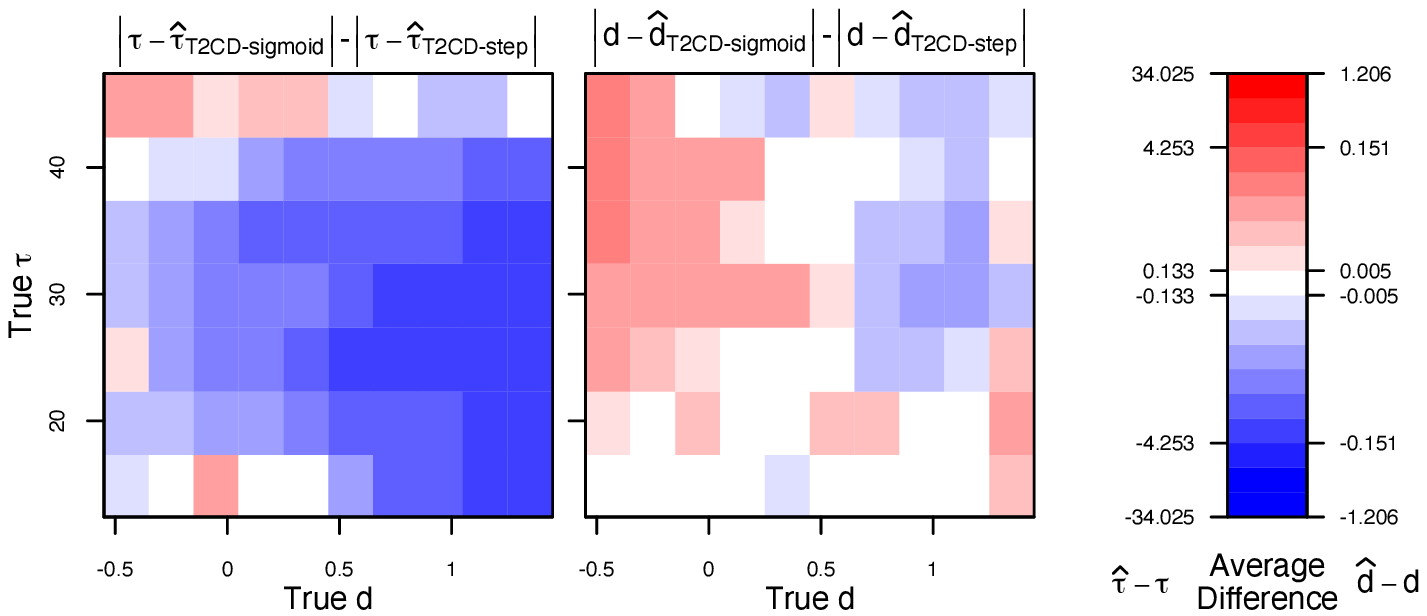}
\caption{Performance of estimates for change location $\tau$ and long-memory parameter $d$ across 100 simulated series per each combination of $\tau$ and $d$. The first regime generated via Gaussian process with squared exponential kernel; the second regime generated via ARFIMA(1,$d$,1).}
\label{fig: univgp_arfima_ea}
\end{figure}

\begin{figure}
\includegraphics{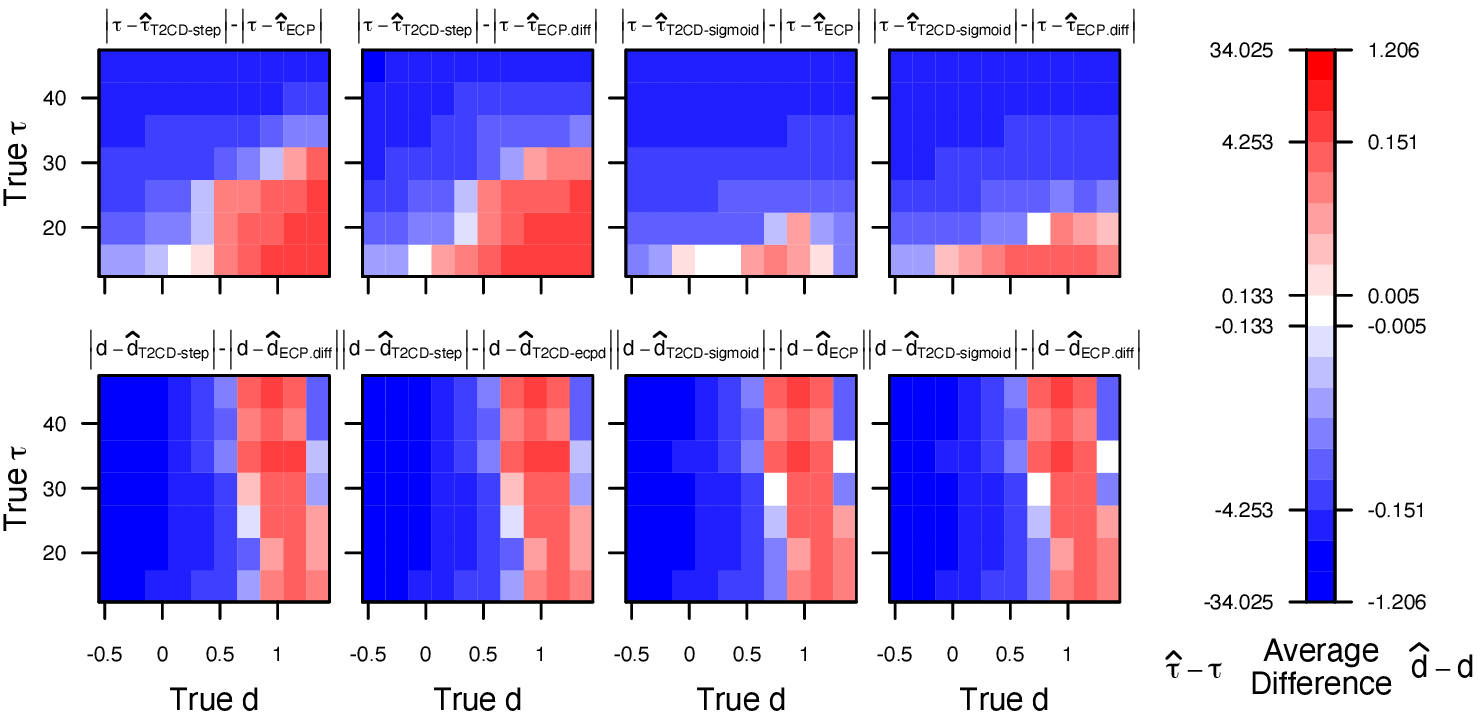}
\caption{Performance of estimates for change location $\tau$ and long-memory parameter $d$ across 100 simulated series per each combination of $\tau$ and $d$. The first regime generated via Gaussian process with squared exponential kernel; the second regime generated via ARFIMA(1,$d$,1).}
\label{fig: univgp_arfima_ecp}
\end{figure}

To assess robustness of change point estimation to deviations from the assumed FI model in the equilibrium period, we conduct a second simulation study in which we simulate equilibrium measurements from an autoregressive moving average fractionally differenced (ARFIMA) model which generalizes the FI model \citep{baillie1996}. We simulate equilibrium regime measurements according to  $(1 - \phi B)(1 - B)^d y_t = (1 + \theta B)\epsilon_t$, where  $\phi,\theta \sim \text{Unif}(0,1)$ and $\epsilon_t \sim \text{N}(0, 0.25)$.  Figures~\ref{fig: univgp_arfima_eaft}-\ref{fig: univgp_arfima_ecp} shows the corresponding estimates of $\tau$ and $d$ obtained using T2CD-step, T2CD-sigmoid, ECP, ECP.diff, FixedTau, and TrueTau.

The relative performance of different estimators when the second regime is generated via ARFIMA(1,$d$,1) is similar to the relative performance of different estimators when the second regime is generated via FI($d$).
This suggests that misspecification of the second regime model does not negate the benefits of using T2CD-step or T2CD-sigmoid.
We continue to observe better estimates of both the change point $\tau$ and the long-memory parameter $d$ from T2CD-sigmoid when the equilibrium process is non-stationary with $d \geq 0.5$.
We also continue to observe better estimates of both the change point $\tau$ and the long-memory parameter $d$ from T2CD-step and T2CD-sigmoid compared to ECP and ECP.diff, with one exception. ECP and ECP.diff sometimes provide better estimates of the change point and long memory parameters $\tau$ and $d$ than T2CD-step and T2CD-sigmoid when $d$ is close to $1$.
Lastly, we continue to observe that T2CD-step and T2CD-sigmoid provided better estimates of the change point $\tau$ than FixedTau, worse estimates of the change point $\tau$ than TrueTau, and comparable estimates of the long memory parameter $d$ relative to FixedTau and TrueTau.

\subsection{Multivariate scenario}

\begin{figure}
\centering
\includegraphics[width=0.6\linewidth]{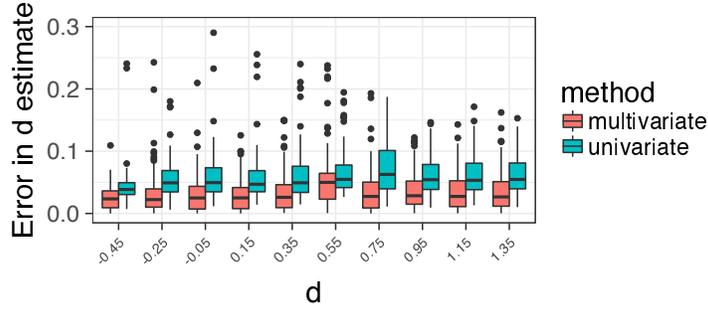}
\caption{Absolute errors in estimates of $d$ by univariate and multivariate implementations of T2CD-step. The pooled estimates reduced errors.}
\label{fig: d_error_mv}
\end{figure}

We compare the absolute errors in the estimates of $d$ by the univariate and multivariate implementations of T2CD-step in Figure \ref{fig: d_error_mv}, which demonstrates the benefit of the pooled estimates in reducing errors across all values of $d$ tested.

\subsection{Discussion}

\begin{figure}
\centering
\includegraphics[width=0.6\linewidth]{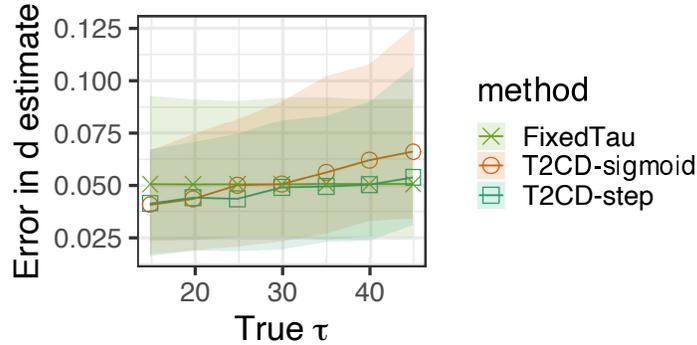}
\caption{Absolute errors in estimates of $d$ for simulation setup where the first regime is generated via Gaussian process with squared exponential kernel and the second regime generated via FI($d$). Medians are plotted and error bars indicate the upper and lower quartiles. T2CD reduced errors when the ground truth $\tau$ is small. For larger $\tau$, T2CD may underestimate $\tau$, which increases errors in estimating $d$.}
\label{fig: d_error}
\end{figure}

We compare the estimates of $d$ by T2CD and the FixedTau method to explore benefits of segmenting the sequence on long-memory parameter estimation. FixedTau segments all sequences at 50. Figure \ref{fig: d_error} plots the absolute errors in the $d$ estimates. Compared to FixedTau, the T2CD methods have lower errors when the ground truth $\tau$ is small. In particular, the upper quartile of T2CD-step error becomes higher than that of FixedTau error only at $\tau = 45$, which is near the upper end of the candidate change point range $\tau_b$. At high values of $\tau$, the T2CD methods may start to segment earlier than the ground truth, causing the observations used for long-memory parameter estimation to be contaminated with first regime observations. Between the two T2CD methods, T2CD-sigmoid show this effect earlier since phase transition is modeled with a smooth curve.

\section{Application to ECIS Data}

\subsection{MDCK cell line}

Table \ref{tab: mdck_d} summarizes the average $d$ estimated by T2CD-step and T2CD-sigmoid across experiments, serum types and infection status. Except for Experiment 2, the average $d$ for infected samples is always higher than that of normal samples.

\begin{table}[]
\centering
\begin{tabular}{lllllllllll}
\hline
\multirow{2}{*}{Expt}                      & \multirow{2}{*}{Serum} & \multirow{2}{*}{Infection} & \multicolumn{2}{l}{T2CD-step} & \multicolumn{2}{l}{T2CD-sigmoid} \\ \cline{4-7} 
                                           &                        &                            & Mean         & SD          & Mean         & SD         \\ \hline
\multicolumn{1}{c}{\multirow{4}{*}{1}}     & \multirow{2}{*}{BSA}   & No                         & 0.730        & 0.063       & 0.728        & 0.073      \\ \cline{3-7} 
\multicolumn{1}{c}{}                       &                        & Yes                        & 0.998        & 0.091       & 1.019        & 0.092      \\ \cline{2-7} 
\multicolumn{1}{c}{}                       & \multirow{2}{*}{Gel}   & No                         & 0.781        & 0.056       & 0.766        & 0.070      \\ \cline{3-7} 
\multicolumn{1}{c}{}                       &                        & Yes                        & 0.873        & 0.157       & 0.943        & 0.101      \\ \hline
\multicolumn{1}{c}{\multirow{4}{*}{2}}     & \multirow{2}{*}{BSA}   & No                         & 0.686        & 0.122       & 0.673        & 0.114      \\ \cline{3-7} 
                                           &                        & Yes                        & 0.656        & 0.109       & 0.824        & 0.099      \\ \cline{2-7} 
                                           & \multirow{2}{*}{Gel}   & No                         & 0.821        & 0.051       & 0.830        & 0.029      \\ \cline{3-7} 
                                           &                        & Yes                        & 0.709        & 0.127       & 0.789        & 0.127      \\ \hline
\multicolumn{1}{c}{\multirow{4}{*}{3}}     & \multirow{2}{*}{BSA}   & No                         & 0.676        & 0.125       & 0.694        & 0.112      \\ \cline{3-7} 
                                           &                        & Yes                        & 0.915        & 0.097       & 0.988        & 0.099      \\ \cline{2-7} 
                                           & \multirow{2}{*}{Gel}   & No                         & 0.757        & 0.118       & 0.782        & 0.097      \\ \cline{3-7} 
                                           &                        & Yes                        & 0.915        & 0.126       & 1.040        & 0.038      \\ \hline
\multicolumn{1}{c}{\multirow{4}{*}{4}}     & \multirow{2}{*}{BSA}   & No                         & 0.676        & 0.110       & 0.702        & 0.072      \\ \cline{3-7} 
                                           &                        & Yes                        & 0.827        & 0.112       & 0.886        & 0.158      \\ \cline{2-7} 
                                           & \multirow{2}{*}{Gel}   & No                         & 0.756        & 0.034       & 0.750        & 0.049      \\ \cline{3-7} 
                                           &                        & Yes                        & 0.805        & 0.186       & 0.834        & 0.160      \\ \hline                                           
\end{tabular}
\caption{MDCK: T2CD estimates of $d$. Average is taken for samples in the same experiment, serum type and infection status.}
\label{tab: mdck_d}
\end{table}

\subsection{BSC cell line}

Table \ref{tab: bsc_d} summarizes the average $d$ estimated by T2CD-step and T2CD-sigmoid across experiments, serum types and infection status. Contrary to the MDCK cell line, the average $d$ for the infected samples in the BSC cell line is lower than that of normal samples in most cases.

\begin{table}[]
\centering
\begin{tabular}{lllllllllll}
\hline
\multirow{2}{*}{Expt}                      & \multirow{2}{*}{Serum} & \multirow{2}{*}{Infection} & \multicolumn{2}{l}{T2CD-step} & \multicolumn{2}{l}{T2CD-sigmoid} \\ \cline{4-7} 
                                           &                        &                            & Mean         & SD          & Mean         & SD         \\ \hline
\multicolumn{1}{c}{\multirow{4}{*}{1}}     & \multirow{2}{*}{BSA}   & No                         & 1.101        & 0.077       & 1.102        & 0.076      \\ \cline{3-7} 
\multicolumn{1}{c}{}                       &                        & Yes                        & 1.003        & 0.056       & 1.017        & 0.044      \\ \cline{2-7} 
\multicolumn{1}{c}{}                       & \multirow{2}{*}{Gel}   & No                         & 1.098        & 0.050       & 1.106        & 0.054      \\ \cline{3-7} 
\multicolumn{1}{c}{}                       &                        & Yes                        & 1.036        & 0.101       & 1.015        & 0.077      \\ \hline
\multicolumn{1}{c}{\multirow{4}{*}{2}}     & \multirow{2}{*}{BSA}   & No                         & 1.057        & 0.092       & 1.057        & 0.083      \\ \cline{3-7} 
                                           &                        & Yes                        & 1.018        & 0.085       & 1.041        & 0.079      \\ \cline{2-7} 
                                           & \multirow{2}{*}{Gel}   & No                         & 1.089        & 0.063       & 1.079        & 0.066      \\ \cline{3-7} 
                                           &                        & Yes                        & 1.021        & 0.069       & 1.012        & 0.047      \\ \hline
\multicolumn{1}{c}{\multirow{4}{*}{3}}     & \multirow{2}{*}{BSA}   & No                         & 1.056        & 0.051       & 1.085        & 0.080      \\ \cline{3-7} 
                                           &                        & Yes                        & 1.079        & 0.063       & 1.072        & 0.053      \\ \cline{2-7} 
                                           & \multirow{2}{*}{Gel}   & No                         & 1.057        & 0.092       & 1.064        & 0.095      \\ \cline{3-7} 
                                           &                        & Yes                        & 1.042        & 0.050       & 1.043        & 0.043      \\ \hline
\multicolumn{1}{c}{\multirow{4}{*}{4}}     & \multirow{2}{*}{BSA}   & No                         & 1.101        & 0.051       & 1.106        & 0.065      \\ \cline{3-7} 
                                           &                        & Yes                        & 1.130        & 0.063       & 1.131        & 0.055      \\ \cline{2-7} 
                                           & \multirow{2}{*}{Gel}   & No                         & 1.114        & 0.060       & 1.119        & 0.045      \\ \cline{3-7} 
                                           &                        & Yes                        & 1.089        & 0.073       & 1.086        & 0.055      \\ \hline                                           
\end{tabular}
\caption{BSC: T2CD estimates of $d$. Average is taken for samples in the same experiment, serum type and infection status.}
\label{tab: bsc_d}
\end{table}

\section{Construction of Original Features}

 The  features described in \citep{gelsinger2017} are
\begin{itemize}
    \item Average resistance at a time fixed time-mark where the measurement tends to peak;
    \item Maximum average resistance;
    \item Average resistance at the end of the sequence;
\end{itemize}
A simple moving average with window length 5 is taken to smoothen the sequence to obtain more stable estimates of the features of interest. 
The time-mark used for the first feature is 17-hour for MDCK and 2-hour for BSC, selected by visual inspection of the data.

\end{appendices}

\end{document}